\documentclass[lettersize,journal]{IEEEtran}

\usepackage{acronym}

\usepackage{cite}
\usepackage{amsmath,amssymb,amsfonts}
\usepackage{algorithmic}
\usepackage{graphicx}
\usepackage{textcomp}
\usepackage{lipsum}
\usepackage[utf8]{inputenc}
\usepackage{textcomp}
\usepackage{caption,subcaption}
\usepackage{cite}
\usepackage{authblk}
\usepackage{amsmath}
\usepackage{nomencl}
\usepackage{enumitem}
\usepackage{epsfig}
\usepackage[table,xcdraw]{xcolor}
\usepackage{booktabs}
\usepackage{upgreek}
\usepackage[normalem]{ulem}
\useunder{\uline}{\ul}{}
\usepackage{longtable}
\usepackage{array}
\usepackage{tabularx}
\usepackage{xcolor}

\usepackage{rotating,multirow,multicol}
\usepackage{wasysym}
\usepackage{adjustbox} 
\usepackage{placeins}
\usepackage{cuted}
\usepackage[linesnumbered,ruled,vlined]{algorithm2e}
\usepackage{commath}
\usepackage{lscape}

\usepackage{amsfonts}

\usepackage{lipsum}
\usepackage{cuted} 

\usepackage{physics}

\usepackage{tikz} 

\acrodef{2D}{two-dimensional}
\acrodef{3D}{three-dimensional}
\acrodef{3GPP}{3rd Generation Partnership Project}
\acrodef{2G}{second generation}
\acrodef{3G}{third generation}
\acrodef{4G}{fourth generation}
\acrodef{5G}{fifth generation}
\acrodef{6G}{sixth generation}

\acrodef{AC}{alternating current}
\acrodef{ACF}{autocorrelation function}
\acrodef{AGC}{automatic gain control}
\acrodef{ACI}{adjacent channel interference}
\acrodef{ACK}{acknowledgement}
\acrodef{AcR}{autocorrelation receiver}
\acrodef{ADC}{analog-to-digital converter}
\acrodef{AF}{amplify-and-forward}
\acrodef{AFL}{anchor-free localization}
\acrodef{AGNSS}{assisted GNSS}
\acrodef{AGPS}{assisted GPS}
\acrodef{AI}{artificial intelligence}
\acrodef{AIC}{Akaike information criterion}
\acrodef{AO}{alternating optimization}
\acrodef{AoA}{angle of arrival}
\acrodef{AoD}{angle of departure}
\acrodef{AOT}{approximate optimum threshold}
\acrodef{AP}{access point}
\acrodef{API}{application programming interface}
\acrodef{ASK}{amplitude shift keying}
\acrodef{ASNR}{accumulated signal-to-noise ratio}
\acrodef{AUB}{asymptotic union bound}
\acrodef{AWGN}{additive white Gaussian noise}

\acrodef{BAN}{body area network}
\acrodef{BAV}{balanced antipodal Vivaldi}
\acrodef{BCH}{Bose--Chaudhuri--Hocquenghem}
\acrodef{BEP}{bit error probability}
\acrodef{BER}{bit error rate}
\acrodef{BF}{brute force}
\acrodef{BFC}{block fading channel}
\acrodef{BIC}{Bayesian information criterion}
\acrodef{BLUE}{best linear unbiased estimator}
\acrodef{BPAM}{binary pulse amplitude modulation}
\acrodef{BPF}{bandpass filter}
\acrodef{BPPM}{binary pulse position modulation}
\acrodef{bps}{bits per second}
\acrodef{BPSK}{binary phase shift keying}
\acrodef{BPZF}{band-pass zonal filter}
\acrodef{BS}{base station}
\acrodef{BSC}{binary symmetric channel}
\acrodef{BTB}{Bellini--Tartara bound}
\acrodef{BD-RIS}{beyond-diagonal RIS}

\acrodef{CAGR}{compound annual growth rate}
\acrodef{CCDF}{complementary cumulative distribution function}
\acrodef{CDF}{cumulative distribution function}
\acrodef{CAD}{computer-aided design}
\acrodef{CAIC}{consistent Akaike information criterion}
\acrodef{CAP}{continuous aperture phased}
\acrodef{CATR}{compact antenna test range}
\acrodef{CCF}{cross-correlation function}
\acrodef{CCI}{co-channel interference}
\acrodef{CI}{close-in}
\acrodef{CIF}{close-in frequency}
\acrodef{CD}{cooperative diversity}
\acrodef{CDMA}{code division multiple access}
\acrodef{CEOT}{channel ensemble optimum threshold}
\acrodef{CEP}{codeword error probability}
\acrodef{CFAR}{constant false alarm rate}
\acrodef{CHF}{characteristic function}
\acrodef{CH}{cluster head}
\acrodef{CIR}{channel impulse response}
\acrodef{CL}{centroid localisation}
\acrodef{CM}{channel model}
\acrodef{CNR}{clutter-to-noise ratio}
\acrodef{CP}{cyclic prefix}
\acrodef{CPR}{channel pulse response}
\acrodef{CR}{channel response}
\acrodef{CRB}{Cramér--Rao bound}
\acrodef{CRC}{cyclic redundancy check}
\acrodef{CLI}{command line interface}
\acrodef{CRLB}{Cramér--Rao lower bound}
\acrodef{CS}{clock skew}
\acrodef{CSCG}{circularly symmetric complex Gaussian}
\acrodef{CSI}{channel state information}
\acrodef{CSMA}{carrier sense multiple access}
\acrodef{CSS}{chirp spread spectrum}
\acrodef{CTS}{clear-to-send}
\acrodef{CW}{continuous wave}
\acrodef{cmWave}{centimetre-wave}

\acrodef{DAA}{detect and avoid}
\acrodef{DAC}{digital-to-analog converter}
\acrodef{DAB}{digital audio broadcasting}
\acrodef{DBB}{digital baseband}
\acrodef{DBPSK}{differential binary phase shift keying}
\acrodef{DBSCAN}{density-based spatial clustering of applications with noise}
\acrodef{DC}{direct current}
\acrodef{DCM}{dual-carrier modulation}
\acrodef{DDP}{detected direct path}
\acrodef{DF}{detect-and-forward}
\acrodef{DFT}{discrete Fourier transform}
\acrodef{DFMS}{monopole dual-feed stripline antenna}
\acrodef{DGPS}{differential GPS}
\acrodef{DLL}{delay-locked loop}
\acrodef{DoA}{direction of arrival}
\acrodef{DoD}{Department of Defense}
\acrodef{DoF}{degrees of freedom}
\acrodef{DP}{direct path}
\acrodef{DR}{detection rate}
\acrodef{DRT}{distance ratio test}
\acrodef{DSSS}{direct-sequence spread spectrum}
\acrodef{DS}{delay spread}
\acrodef{DTR}{differential transmitted-reference}
\acrodef{DTT}{diffraction tomography theory}
\acrodef{DVBH}{digital video broadcasting--handheld}
\acrodef{DVBT}{digital video broadcasting--terrestrial}
\acrodef{DUT}{device under test}

\acrodef{EH}{energy harvesting}
\acrodef{EM}{electromagnetic}
\acrodef{EC}{European Commission}
\acrodef{ED}{energy detector}
\acrodef{EE}{energy efficiency}
\acrodef{EMF}{electromagnetic field}
\acrodef{EDR}{energy detector receiver}
\acrodef{EFIM}{equivalent Fisher information matrix}
\acrodef{EIRP}{effective isotropic radiated power}
\acrodef{EKF}{extended Kalman filter}
\acrodef{EMF}{electromagnetic fields}
\acrodef{KKT}{Karush--Kuhn--Tucker}
\acrodef{ELP}{equivalent low-pass}
\acrodef{EMCB}{extended Miller--Chang bound}
\acrodef{EME}{minimum eigenvalue ratio detector}
\acrodef{EMI}{electromagnetic interference}
\acrodef{ENP}{estimated noise power}
\acrodef{ESA}{European Space Agency}
\acrodef{EU}{European Union}
\acrodef{EVD}{eigenvalue decomposition}
\acrodef{EVM}{error vector magnitude}
\acrodef{EEPROM}{electrically erasable programmable read-only memory}
\acrodef{EMBG}{electromagnetic band gap}
\acrodef{MF-RIS}{multi-functional RIS}
\acrodef{MPPT}{maximum power point tracking}

\acrodef{RMS}{root mean square}

\acrodef{FAR}{false alarm rate}
\acrodef{FCC}{Federal Communications Commission}
\acrodef{FDMA}{frequency division multiple access}
\acrodef{FEC}{forward error correction}
\acrodef{FF}{far-field}
\acrodef{FR}{frequency range}
\acrodef{FFT}{fast Fourier transform}
\acrodef{FG}{factor graph}
\acrodef{FH}{frequency hopping}
\acrodef{FMCW}{frequency-modulated continuous wave}
\acrodef{FPGA}{field-programmable gate array}
\acrodef{FIM}{Fisher information matrix}
\acrodef{FLL}{frequency-locked loop}
\acrodef{FS}{frame synchronization}
\acrodef{FT}{Fourier transform}
\acrodef{FOV}{field of view}

\acrodef{GA}{Gaussian approximation}
\acrodef{GD}{gradient descent}
\acrodef{GDOP}{geometric dilution of precision}
\acrodef{GLR}{generalized likelihood ratio}
\acrodef{GLRT}{generalized likelihood ratio test}
\acrodef{GML}{generalized maximum likelihood}
\acrodef{GPRS}{general packet radio service}
\acrodef{GBSM}{geometry-based stochastic model}
\acrodef{GPS}{global positioning system}
\acrodef{GPIO}{general-purpose input/output}
\acrodef{GUI}{graphical user interface}
\acrodef{ETSI}{European Telecommunications Standards Institute}
\acrodef{GR}{group report}
\acrodef{GS}{group specification}
\acrodef{PTS}{power-time splitting}
\acrodef{DRAM}{dynamic random memory access}
\acrodef{MCU}{microcontroller}

\acrodef{HAP}{high-altitude platform}
\acrodef{HCRB}{hybrid Cramér--Rao bound}
\acrodef{HMM}{hidden Markov model}
\acrodef{HPA}{high-power amplifier}
\acrodef{HPBW}{half-power beamwidth}
\acrodef{PRRP}{peak re-radiated power}
\acrodef{PEB}{position error bound}
\acrodef{PLE}{path loss exponent}
\acrodef{RA}{reflection amplifier}
\acrodef{SWIPT}{simultaneous wireless information and power transfer}

\acrodef{ICT}{information and communication technologies}
\acrodef{ICS}{Institute for Communication Systems}
\acrodef{IEEE}{Institute of Electrical and Electronics Engineers}
\acrodef{I$^2$C}{inter-integrated circuit}
\acrodef{IF}{intermediate frequency}
\acrodef{IP}{ingress protection}
\acrodef{IFFT}{inverse fast Fourier transform}
\acrodef{IMU}{inertial measurement unit}
\acrodef{INR}{interference-to-noise ratio}
\acrodef{INS}{inertial navigation system}
\acrodef{ISG}{Industry Specification Group}
\acrodef{ITU}{International Telecommunication Union}
\acrodef{IoT}{internet of things}
\acrodef{IIoT}{industrial internet of things}
\acrodef{IR}{impulse radio}
\acrodef{IRS}{intelligent reflecting surface}
\acrodef{ISAC}{integrated sensing and communication}
\acrodef{IC}{integrated circuit}
\acrodef{ISI}{inter-symbol interference}
\acrodef{FoV}{field of view}

\acrodef{KF}{Kalman filter}
\acrodef{KPI}{key performance indicator}

\acrodef{LDPC}{low-density parity-check}
\acrodef{LIS}{large intelligent surface}
\acrodef{LLR}{log-likelihood ratio}
\acrodef{LRT}{likelihood ratio test}
\acrodef{LNA}{low-noise amplifier}
\acrodef{LoS}{line of sight}
\acrodef{MPC}{multipath component}
\acrodef{PAM}{programmable active matrix}
\acrodef{PIN}{positive-intrinsic-negative}
\acrodef{PDP}{power delay profile}
\acrodef{PEC}{perfect electric conductor}

\acrodef{QPSK}{quadrature phase shift keying}

\acrodef{FET}{field effect transistor}
\acrodef{BJT}{bipolar junction transistor}
\acrodef{MIMO}{multiple-input multiple-output}
\acrodef{SU-MIMO}{single user multiple-input multiple-output}
\acrodef{ML}{machine learning}
\acrodef{mmWave}{millimeter wave}
\acrodef{MMSE}{minimum mean-square error}
\acrodef{MPC}{multipath component}
\acrodef{MJ}{multi-junction}
\acrodef{MRI}{magnetic resonance imaging}
\acrodef{NF}{near-field}
\acrodef{NR}{new radio}
\acrodef{NLoS}{non-line-of-sight}
\acrodef{NGA}{Next G Alliance}
\acrodef{NPL}{National Physical Laboratory}
\acrodef{NOMA}{non-orthogonal multiple acces}
\acrodef{OFDM}{orthogonal frequency division multiplexing}
\acrodef{OOK}{on-off keying}
\acrodef{OTA}{over-the-air}
\acrodef{OAM}{orbital angular momentum}
\acrodef{PoC}{proof of concept}
\acrodef{RISTA}{RIS Tech Alliance}
\acrodef{RTO}{real-time oscilloscope}

\acrodef{S11}{scattering parameter 11}
\acrodef{S21}{scattering parameters 21}
\acrodef{SDO}{standard development organization}
\acrodef{SBC}{single-board computer}
\acrodef{SPI}{serial peripheral interface}
\acrodef{SoC}{system on chip}

\acrodef{ToF}{time of flight}
\acrodef{TDD}{time division duplexing}
\acrodef{TR}{technical report}
\acrodef{TS}{technical specification}
\acrodef{TE}{transverse electric}
\acrodef{TM}{transverse magnetic}
\acrodef{TRRP}{total re-radiated power}
\acrodef{LiDAR}{light detection and ranging}
\acrodef{mMIMO}{massive multiple-input multiple-output}
\acrodef{MEC}{mobile edge computing}

\acrodef{PA}{power amplifier}
\acrodef{PAM}{programmable active matrix}
\acrodef{PDF}{probability density function}
\acrodef{PHY}{physical layer}
\acrodef{PLL}{phase-locked loop}
\acrodef{PCB}{printed circuit board}
\acrodef{PSK}{phase shift keying}
\acrodef{TX}{transmitter}
\acrodef{RX}{receiver}
\acrodef{RCS}{radar cross section}
\acrodef{RCC}{radar and communications coexistence}
\acrodef{SC}{solar cell}
\acrodef{SIM}{stacked intelligent metasurface}
\acrodef{SWE}{smart wireless environment}

\acrodef{QAM}{quadrature amplitude modulation}
\acrodef{QoS}{quality of service}

\acrodef{PPE}{personal protective equipment}

\acrodef{RADAR}{radar}
\acrodef{RF}{radio frequency}
\acrodef{RFID}{radio-frequency identification}
\acrodef{RIS}{reconfigurable intelligent surface}
\acrodef{RSSI}{received signal strength indicator}
\acrodef{RSRP}{reference signal received power}
\acrodef{RMSE}{root-mean-square error}
\acrodef{ROC}{receiver operating characteristic}
\acrodef{O-RAN}{open radio access network}
\acrodef{PC}{Polycarbonate}

\acrodef{SAR}{specific absorption rate}
\acrodef{SISO}{single-input single-output}
\acrodef{SLAM}{simultaneous localization and mapping}
\acrodef{SNR}{signal-to-noise ratio}
\acrodef{STAR}{simultaneously transmitting and reflecting}
\acrodef{SRR}{split ring resonator}
\acrodef{SLoS}{sidelobe line of sight}
\acrodef{SMA}{subminiature version A}
\acrodef{SV}{Saleh-Valenzuela}

\acrodef{TDMA}{time division multiple access}
\acrodef{TDOA}{time difference of arrival}
\acrodef{THz}{terahertz}
\acrodef{TOA}{time of arrival}
\acrodef{TI}{Texas Instruments}

\acrodef{UAV}{unmanned aerial vehicle}
\acrodef{UE}{user equipment}
\acrodef{UWB}{ultra-wideband}

\acrodef{V2X}{vehicle-to-everything}
\acrodef{VR}{virtual reality}
\acrodef{vLoS}{virtual line of sight}
\acrodef{VNA}{vector network analyser}

\acrodef{WDC}{wave domain computing}
\acrodef{WLAN}{wireless local area network}
\acrodef{WSN}{wireless sensor network}
\acrodef{WPT}{wireless power transfer}
\acrodef{WI}{work item}

\acrodef{XPR}{cross polarisation power ratio}

\acrodef{ZoA}{zenith angle of arrival}
\acrodef{ZoD}{zenith angle of departure}

\makeatletter
\newcommand{\fs@ruled}{%
  \def\@fs@cfont{\normalfont\scriptsize\bfseries}%
  \let\@fs@capt\floatc@ruled
  \def\@fs@pre{\hrule height.8pt depth0pt \kern2pt}%
  \def\@fs@post{\kern2pt\hrule\relax}%
  \def\@fs@mid{\kern2pt\hrule\kern2pt}%
  \let\@fs@iftopcapt\iftrue}
\makeatother

\normalsize

\begin{document}
%

\title{Multi-Functional RIS-enabled Radar and Communication Coexistence: Channel Modeling and a Sub-6~GHz Indoor Measurement Campaign}


\author{Anton~Tishchenko,\thanks{\textit{Anton~Tishchenko, Demos~Serghiou, Hamidreza~Taghvaee, Ahmed~Elzanaty, Gabriele~Gradoni, Mohsen~Khalily, and Rahim Tafazolli are with 5G $\&$ 6G Innovation center (5GIC $\&$ 6GIC), Institute for Communication Systems (ICS), University of Surrey, Guildford, GU2 7XH, U.K. (email:$\{$a.tishchenko, demos.serghiou, h.taghvaee, a.elzanaty, m.khalily, r.tafazolli$\}$@surrey.ac.uk).}}~\IEEEmembership{Student Member,~IEEE}, Demos~Serghiou,~\IEEEmembership{Member,~IEEE}, Hamidreza~Taghvaee,~\IEEEmembership{Member,~IEEE}, Arman~Shojaeifard,\thanks{\textit{Arman Shojaeifard is with InterDigital, London EC2A 3QR, U.K. (email: \\arman.shojaeifard@interdigital.com).}}~\IEEEmembership{Senior Member, IEEE}, Ahmed~Elzanaty,~\IEEEmembership{Senior~Member,~IEEE}, Gabriele~Gradoni,~\IEEEmembership{Member,~IEEE},
Fraser~Burton,\thanks{\textit{Fraser Burton is with Applied Research, British Telecom, EC1A 7AJ, London, U.K. (email: fraser.burton@bt.com).}} Mohsen~Khalily,~\IEEEmembership{Senior~Member,~IEEE}, and~Rahim~Tafazolli,~\IEEEmembership{Fellow, IEEE}
\thanks{The work of Anton~Tishchenko was supported in part by the U.K. Engineering and Physical Sciences Research Council (EPSRC) iCASE studentship with British Telecom under Grant EP/W522272/1.}
}

\markboth{(For Review)}%
{Submitted paper}

\maketitle

\begin{abstract}
In this work, we analyze a multi-functional reconfigurable intelligent surface (MF-RIS)-enabled radar and communication coexistence (RCC) system, detailing the key aspects of its phase synthesis codebook generation and the implemented localization algorithm for real-time user tracking based on density-based spatial clustering of applications with noise (DBSCAN), which features a Kalman filter for the prediction of user mobility. We derived a 3GPP-compatible radar cross-section (RCS) and re-radiation pattern-based channel model for the described MF-RIS system, supplementing it with channel measurements. We obtained large and small-scale characteristics, including path loss, shadow fading, Rician K-factor, cluster powers, and RMS delay spread.
The study finds that Sub-6 GHz indoor propagation is largely free of blind spots, even with a blocked line-of-sight (LoS) path. Therefore, the proposed channel model includes non-line-of-sight (NLoS) paths, including the ones created by the MF-RIS. We also performed an experimental evaluation of the channel throughput in a fifth generation (5G) new radio (NR) single user multiple-input-multiple-output (SU-MIMO) system, reporting a 74\% reduction in throughput variance and a 12.5\% sum-rate improvement within the MF-RIS near-field compared to the no-RIS setup. This result shows that the MF-RIS can minimize delay spread and increase the coherence bandwidth by creating virtual-LoS (vLoS) path for the moving user, thereby effectively hardening wireless MIMO channels.

\end{abstract}

\begin{IEEEkeywords}
5G NR, beam focusing, channel hardening, channel measurement, delay spread, localization, multi-functional RIS, near-field, power delay profile, Sub-6~GHz, SU-MIMO.
\end{IEEEkeywords}

\IEEEpeerreviewmaketitle

\section{Introduction}

\IEEEPARstart{C}{ellular} wireless networks are continuously evolving, as the private \ac{5G} network sector is projected to grow at a \ac{CAGR} of 40.5\% between 2024 and 2032 \cite{Markets}. Such rapid growth places unprecedented demands on the \ac{5G} network quality and power consumption, which has been the cornerstone issue surrounding \ac{5G} deployments to date. The emergence of \ac{RIS} technology \cite{RIS1, RIS2, RIS3} promised to address this problem, by adding smart surfaces in the wireless propagation environment that consume negligible amounts of power, when compared to base stations or active relays \cite{Relays}. However, this promise remains largely unfulfilled due to several critical factors concerning \ac{RIS} deployments that make such structures unattractive for network vendors, such as their costs \cite{FoM}, possibility of causing interference to other network equipment \cite{BoW}, and deployment complexities \cite{BT}. Despite this negative outlook, the recent emergence of \ac{ISAC} as the key technology for the \ac{6G} of wireless networks \cite{ISAC1} is likely to cause a renewed interest in the \ac{RIS} technology, since it has a potential to serve as the platform for enabling the so-called `loose integration level' of \ac{ISAC} \cite{ISAC2} with minimal hardware or software changes in the \ac{5G} network architecture. \par 
In this light, the emergence of \ac{MF-RIS} systems with advanced capabilities, such as \ac{UE} localization and \ac{EH}, has recently been envisioned in the literature \cite{MF-RIS1, MF-RIS2, MF-RIS3, MF-RIS4}. These systems are projected to be almost entirely independent of \ac{BS} processing and capable of unrivaled energy efficiency and environmental awareness through the integration of additional technologies, such as radar\cite{Radar}, solar panels \cite{Solar}, or through stacked multi-layer designs capable of performing wave-domain computations on the \ac{RIS} itself\cite{SIM}. \par
However, the literature lacks detailed studies of practical \ac{MF-RIS} systems and localization algorithms, in particular concerning channel modeling and near-field measurements, which dominates the majority of indoor deployment scenarios due to the size of \ac{MF-RIS} being relatively large compared to the wavelength of carrier signals. 

\begin{table*}[!htp]
\centering
    \captionsetup{justification=centering}\captionsetup{ labelfont={sc,footnotesize}}
\caption{\footnotesize\sc { Summary of RIS Channel Measurements in the Prior Art}}
\label{tb:intro}
\renewcommand*{\arraystretch}{1.25}

\begin{tabular}{|p{2.5cm}|p{1.5cm}|p{1cm}|p{1.5cm}|p{1.5cm}|p{2.5cm}|p{2.5cm}|p{1.5cm}|}
\hline

\textbf{Measurement Scenario}
& \textbf{Operating Freq.}   & \textbf{Phase Res.}    
& \textbf{Evaluated distance} 
& \textbf{\ac{RIS} number of unit cells}
& \textbf{Applied channel model}                             
& \textbf{Localization / Sensing}   
& \textbf{Ref.} 
\\ \hline
Anechoic chamber
& 4.25~GHz and 10.25~GHz
& 1-bit
& Far-field
& 8~$\times$~32, 50~$\times$~34, and 100~$\times$~102
& Path loss model
& Not considered
& \cite{bMes1}
\\ \hline
Anechoic chamber
& 27~GHz and 33~GHz
& 1-bit
& Far-field
& 20~$\times$~56 and 40~$\times$~40
& Path loss model
& Not considered
& \cite{bMes2}
\\ \hline
Indoor and outdoor 
& 2.6~GHz, 4.9~GHz, and 26~GHz
& 1-bit
& Far-field
& 20~$\times$~20, 32~$\times$~32, and 64~$\times$~64
& Measurement only
& Not considered
& \cite{bMes3}
\\ \hline
Indoor and outdoor
& 5.8~GHz
& 2-bit
& Far-field
& 55~$\times$~20
& Measurement only
& Greedy~fast beamforming~algorithm
& \cite{bMes13}
\\ \hline
Indoor
& 28~GHz
& 1-bit
& Far-field
& 20~$\times$~20 
& Measurement only
& OFDM-based ISAC
& \cite{bMes4}
\\ \hline
Indoor and outdoor
& 2.6~GHz
& 1-bit
& Far-field
& 32~$\times$~16
& Path loss model
& Not considered
& \cite{bMes5} 
\\ \hline
Indoor and outdoor
& 2.6~GHz
& 1-bit
& Far-field
& 32~$\times$~16
& Path loss model
& Not considered
& \cite{bMes55} 
\\ \hline
Indoor and outdoor
& 2.6~GHz
& 1-bit
& Far-field
& 32~$\times$~16
& Time-variant cascaded GBSM
& Not considered
& \cite{bMes6} 
\\ \hline
Outdoor
& 5.8~GHz
& 3-bit
& Far-field
& 16~$\times$~10
& Effective baseband equivalent model
& Beam scanning
& \cite{bMes7}
\\ \hline
Indoor
& 27~GHz
& 1-bit
& Far-field
& 40~$\times$~40
& Measurement only
& Not considered
& \cite{bMes8}
\\ \hline
Indoor
& 3.75~GHz
& 3-bit
& Far-field
& 48~$\times$~48
& Measurement only
& Not considered
& \cite{bMes99}
\\ \hline
Indoor
& 26.9~GHz
& 2-bit
& Far-field
& 45~$\times$~50
& Measurement only
& Not considered
& \cite{bMes10}
\\ \hline
Indoor
& 26~GHz
& 3-bit
& Far-field
& 96~$\times$~96
& Raytracing
& Not considered
& \cite{bMes12}
\\ \hline
Indoor
& 3.5~GHz
& 4-bit
& Near-field and far-field
& 50~$\times$~37
& Time-variant cascaded GBSM
& Radar and communication coexistence
& \textbf{This work}
\\ \hline

\end{tabular}

\end{table*}

\subsection{Related Work and Motivation}
The summary of related state-of-the-art \ac{RIS} channel measurement works is provided in Table \ref{tb:intro}, with a particular focus on \ac{UE} localization methods and near-field measurement. The literature review outlines several key findings:
\begin{itemize}
    \item \ac{RIS}-assisted channel propagation in the near-field remains largely uncharacterized with only several works in the literature that characterize the \ac{RIS} near-field under free-space conditions and delve into its near-field codebook design in \cite{RIS-NF1,RIS-NF2, RIS-NF3}.
    \item Similarly, \ac{RIS}-assisted \ac{UE} localization remains largely unexplored through measurement, with the majority of works dedicated to triangulation-based methods for \ac{UE} localization, such as \cite{Coloris, bWankai, bAhmed} - a concept that can be referred to as `opportunistic \ac{ISAC}'. On the other hand, active \ac{RIS}-based localization through the means of radar integration has only been theorized by the authors of \cite{Alkhateeb1} and \cite{SensingRIS}, who considered the addition of active sensors to the \ac{RIS}. Also, the authors of \cite{bAshwin, bAliAli} proposed \ac{OAM}-based \ac{MF-RIS} with radar that is also able to perform \ac{ISAC} tasks at the same frequency without mutual interference.
    \item We can also observe that the literature lacks detailed studies of the fast-fading channel models of the \ac{RIS}-assisted communication, with only a few examples in \cite{bMes6, GBSM} that consider the \ac{GBSM} for \ac{RIS}-assisted communication. However, these studies neglect the \ac{NLoS} channel between the \ac{BS} and the \ac{UE}, and therefore cannot be considered accurate for indoor environments with omnipresent \acp{MPC}. Additionally, the authors of \cite{GBSM} utilized large bandwidth and large antenna array simplifications for \ac{RIS}-assisted channel modeling, which is also incorrect, because \ac{RIS} is a scatterer that does not transmit active power. Furthermore, these works were not validated through practical \ac{MIMO} system-level measurements and do not explain how a near-field phase synthesis of a \ac{RIS} affects \acp{PLE} and the fading distribution of the wireless channel. 
\end{itemize} 


\subsection{Contributions}
To address the gaps highlighted in the literature review, this paper presents an indoor channel measurement campaign using an \ac{MF-RIS} with integrated \ac{mmWave} radar for user localization, performed in the \ac{MF-RIS} near-field. The exact contributions are detailed as follows:
\begin{itemize}
    \item We derived a simplified fast-fading channel model for the \ac{MF-RIS}-assisted communication, which is based on \ac{RCS} and re-radiation patterns of the \ac{RIS} for each considered reflection angle. This model is compatible with the \ac{3GPP} models in \cite{3GPP1} and \cite{3GPP2}, paving the way for \ac{MF-RIS} to be specified in ray-tracing simulators as a component with a variable \ac{RCS} and pre-measured re-radiation patterns.
    \item We show that the conventional Rayleigh fading channel model is not suitable for modeling the \ac{MF-RIS}-assisted communications channel, which is described better with the Weibull distribution over a wide bandwidth or the log-normal distribution over a narrow bandwidth. 
    \item We provide a comprehensive evaluation of large-scale and small-scale channel parameters with and without \ac{MF-RIS} in two measurement setups (a large indoor space and a medium-sized conference room). 
    \item We detail the \ac{MF-RIS} codebook phase synthesis methods for the near-field and the far-field regimes, highlighting their differences and providing the exact derivations.
    \item We obtain path loss exponents from empirical measurements and provide histograms for the fast-fading distribution in the BS-RIS-UE channel.
    \item We present a \ac{UE} localization algorithm for the described \ac{MF-RIS} system, based on \ac{DBSCAN} and a Kalman filter applied to each coordinate. We characterize its performance in terms of the average 127~ms prediction latency and the average 6.47$^\circ$ accuracy in \ac{AoA} estimation.
    \item We perform channel throughput measurement on a 2 $\times$ 2 \ac{5G} \ac{NR} \ac{SU-MIMO} system, and show that the addition of \ac{MF-RIS} in the communication channel results in a 74\% reduction in throughput variance and a 12.5\% sum-rate improvement within its near-field.
\end{itemize}

The rest of the paper is organized as follows: Sections \ref{ch2} and \ref{ch3} provide the theoretical background for the analysis of the \ac{MF-RIS}-enabled \ac{RCC} system, where \ac{RCC} tasks are performed independently from each other without interference. Then, Section \ref{ch4} provides the localization algorithm for the \ac{MF-RIS}-enabled \ac{RCC} system and the theoretical background of the \ac{MF-RIS}-induced channel hardening effect. This discussion is further complimented by the derivation of the near-field and far-field codebook phase synthesis in Section \ref{ch5}. Finally, the channel measurement setup and results are provided in Sections \ref{ch6} and \ref{ch7}. All results from the channel measurement campaign are summarized in several tables, and the paper is concluded with the throughput measurement of the 2~$\times$~2 \ac{5G} \ac{NR} \ac{SU-MIMO} system with a dynamically moving \ac{UE}.

\section{Radar Channel Model}\label{ch2}
This section details the theoretical analysis of the \ac{mmWave} \ac{MIMO} radar system considered in this work. 

\subsection{Signal Model}
Consider a radar system equipped with an array comprising $M_{\mathrm{A}}$ transmit antennas and $N_{\mathrm{A}}$ receive antennas. The transmitted signal is expressed as:
\begin{equation}
    \mathbf{x}= \Phi \mathbf{s},
\end{equation}
where $\mathbf{x}=[\mathbf{x}_1,\,\mathbf{x}_2,\,\cdots,\,\mathbf{x}_{M_{\mathrm{A}}}]^{\mathsf{T}}$, $\Phi \in \mathbb{R}^{M_{\mathrm{A}}\times p}$ denotes the precoding matrix, and $\mathbf{s} \in \mathbb{R}^{p\times 1}$ is the signal vector prior to precoding.

The received signal at the $n$th receive antenna, with $n \in \mathsf{N}_{\mathrm{A}} \triangleq \{1,2,\cdots,N_{\mathrm{A}}\}$, is given by:
\begin{equation}\label{eq:rxsignal}
    z_n=\mu_n + w_n,
\end{equation}
where $w_n\sim \mathcal{CN}\!\left( 0, \sigma^2\right)$ is a circularly symmetric complex Gaussian random variable with variance $\sigma^2$, and
\begin{equation}
    \mu_{n}= \kappa \sum_{m=1}^{M_{\mathrm{A}}} x_m 
    \exp\!\left( -j \frac{2 \pi}{\lambda} (d_{mo}+d_{on}) \right).
\end{equation}
Here, $\lambda$ denotes the wavelength, $d_{mo}$ is the distance between the $m$th transmit antenna and the target, and $d_{on}$ is the distance between the target and the $n$th receive antenna.

The channel amplitude is modeled as:
\begin{equation}
    \kappa = e \rho,
\end{equation}
where $\rho$ is the complex scattering coefficient, which depends on the wavelength and the radar cross-section, and $e$ is defined as:
\begin{equation}
    e= \sqrt{\frac{\lambda^2 }{d_{\mathrm{tx}}^{2} (4\pi)^2} \frac{\lambda^2}{d_{\mathrm{rx}}^{2} (4\pi)^2}},
\end{equation}
where $d_{\mathrm{tx}}$ and $d_{\mathrm{rx}}$ denote the distances from the transmit and receive array phase centers to the target, respectively.

Based on the received signal model in \eqref{eq:rxsignal}, the \ac{CRLB} is derived to assess the sensing performance of the \ac{MIMO} radar system.

\subsection{\ac{CRLB} Derivation}
The performance of any unbiased estimator $\hat{\boldsymbol{\theta}}$ can be lower-bounded by the \ac{CRLB}:
\begin{equation}
    \mathsf{CRLB}=\mathrm{tr}\!\left(\mathbf{J}^{-1}\!\left( {\boldsymbol{\theta}}\right)\right),
\end{equation}
where ${\boldsymbol{\theta}}=\left[x_r,y_r\right]^{\mathsf{T}}$ denotes the target location vector and $\mathbf{J}$ is the Fisher information matrix (FIM). The $(i,j)$th element of $\mathbf{J}$ is given by:
\begin{equation}\label{eq:FIM}
\left[\mathbf{J}\!\left( {\boldsymbol{\theta}}\right)\right]_{i,j}
=\frac{2}{\sigma ^2} \operatorname{Re}\!\left\lbrace \sum _{n=1}^{N_{\mathrm{A}}}
\frac{\partial \mu_{n}^{*}}{\partial \left[{\boldsymbol{\theta}}\right]_i}
\frac{\partial \mu_{n}}{\partial \left[{\boldsymbol{\theta}}\right]_j}
\right\rbrace,
\end{equation}
where $\left[\cdot\right]_a$ extracts the $a$th element of its argument and $\left[\cdot\right]_{a,b}$ denotes the $(a,b)$th entry of its argument.


\section{Communication Channel Model}\label{ch3}
In \ac{RIS}-assisted communication, which may be \ac{SISO}, \ac{MIMO}, or anything in between, the \ac{BS}-\ac{RIS}-\ac{UE} channel consists of three distinctive parts that require independent modeling: 
\begin{enumerate}
    \item The \ac{BS}-\ac{RIS} channel, which is the sum of \ac{LoS} paths between the \ac{BS} and the \ac{RIS}.
    \item The \ac{BS}-\ac{UE} channel, which is the sum of \acp{MPC} from \ac{NLoS} paths between the \ac{BS} and the \ac{UE}.
    \item The \ac{RIS}-\ac{UE} channel, which is dominated by the sum of \ac{LoS} paths between the \ac{RIS} and the \ac{UE}. However, it also includes \acp{MPC} created by the \ac{RIS} in the environment, implying additional \ac{NLoS} paths created by the \ac{RIS}.
\end{enumerate}
The combination of \ac{LoS} \ac{BS}-\ac{RIS} and \ac{LoS} \ac{RIS}-\ac{UE} paths is typically referred to as the \ac{vLoS} component of the \ac{RIS}-assisted channel. While the \ac{vLoS} component dominates the near-field of the \ac{RIS}, its effects diminish in the far-field of the \ac{RIS}, where the sum of all \ac{NLoS} paths starts to dominate instead. 
\subsection{Path Loss Model}
\begin{figure}[!t]
\centering

    \begin{turn}{0}
    \includegraphics[width=\columnwidth]{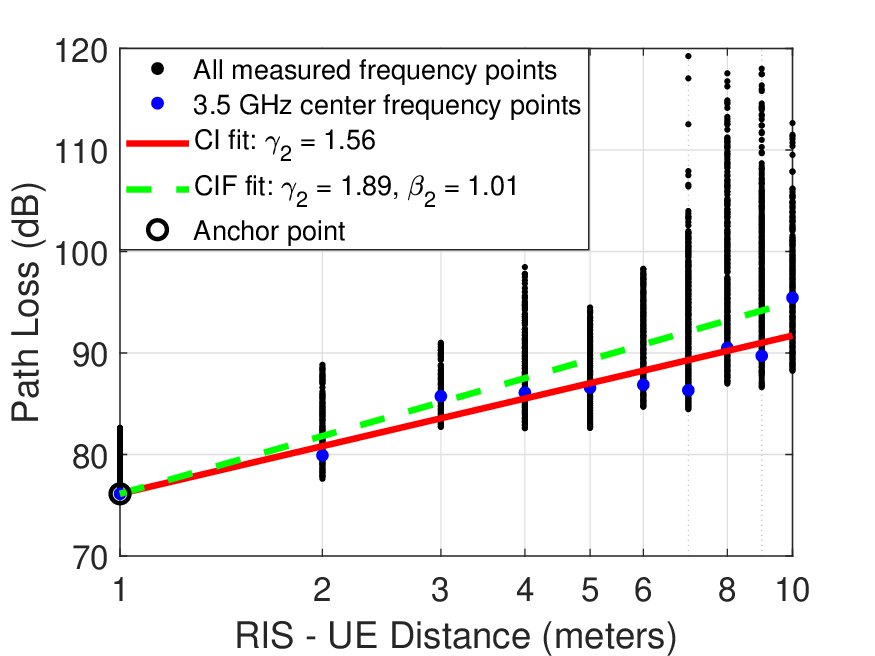} 
    \end{turn}

\caption{Measured $h_{\text{BS-RIS-UE}}$ path loss exponent(s) based on the CI and the CIF model fits with variable $d_{\text{RIS-UE}}$.}
\label{fig:pathlossris}
\end{figure}

To obtain the absolute receive power $\text{P}^{\text{Rx}}_u$ 
at the \ac{UE}, the following expression is applied to the BS-RIS-UE path:
\begin{equation}
 P^{\text{Rx}}_{u}(t)
= \frac{1}{T_1}
\int_{t_0}^{t_0 + T_1}
P^{\text{Tx}}_{s}\;
\frac{\displaystyle\sum_{\ell}\left|h^{\text{BS-RIS-UE}}_{u,s,\ell}(t)\right|^2}
{10^{\text{PL}^{\text{BS-RIS-UE}}_{\text{CI}}/10}}
\, dt,
\end{equation}
\noindent
where $t_0$ is the starting time of the averaging integral, $T_1$ is the duration over which the received power is averaged, $u \times s$ are the antennas on \ac{BS} and \ac{UE} terminals, $\text{P}^{\text{Tx}}$ is the transmit power, $\text{PL}^{\text{BS-RIS-UE}}_{\text{CI}}$ is the \ac{CI} path loss of the BS-RIS-UE path in $\text{dB}$, $h^{\text{BS-RIS-UE}}_{u,s,\ell}(t)$ is the $\ell$-th tap of the fast-fading \ac{CIR}, and $\ell$ indexes the channel delay taps. In this work we reuse \ac{3GPP} \ac{TR} 38.901 conventions, while introducing new, \ac{RIS}-specific definitions. Therefore, we use the \ac{CI} path loss model for the BS-RIS-UE path, expressed as:
\begin{align} \nonumber
    &\text{PL}^{\text{BS-RIS-UE}}_{\text{CI}}(d_{\text{BS-RIS}}, d_{\text{RIS-UE}}) = \\& \nonumber 
    20 \log_{10}\left( \frac{4\pi d_0}{\lambda} \right) + 10 \gamma_1 \log_{10}\left(\frac{ d_{\text{BS-RIS}}}{d_0}\right) \\& + 10 \gamma_2 \log_{10}\left(\frac{ d_{\text{RIS-UE}} }{d_0 }\right) - 10 \log_{10}\left( PL^{\text{FS}}_{\text{RIS}}\right)
     + X^{\text{CI}}_{\sigma},
\end{align}
\noindent
where $d_{\text{BS-RIS}}$ = $\left\|\mathbf{r}_{\text{BS-RIS}}\right\|$, $d_{\text{RIS-UE}}$ = $\left\|\mathbf{r}_{\text{RIS-UE}}\right\|$, $d_0$ is the reference distance (typically set to 1~m), $\lambda$ is the wavelength of the operating frequency, $PL^{\text{FS}}_{\text{RIS}}$ is the free space path loss of the RIS, $\gamma_1$ and $\gamma_2$ are the \acp{PLE} related to the variation of distance\footnotemark[2], and $X^{\text{CI}}_{\sigma_{\text{SF}}} \sim \mathcal{N}(0, \sigma_{\text{SF}}^2)$ is the shadow fading term in dB, which accounts for large-scale power variations (due to blockages, etc.). The measured path loss and the resulting exponents are shown in Fig. \ref{fig:pathlossris}, which shows that the \ac{CI} \ac{PLE} of the RIS-UE channel is less than the conventional $\gamma=2$. 

\footnotetext[2]{The \ac{CI} path loss model can be expanded to include a reference frequency $f_0$ and a frequency exponent $\beta$ terms to obtain the \ac{CIF} model by substituting 10$\gamma \left(1 + \beta \frac{f-f_0}{f_0}\right) \log_{10}\left(\frac{d}{d_0}\right)$ into each distance term. The \ac{CIF} model is considered to be more accurate, yet is less common.} 
The $PL^{\text{FS}}_{\text{RIS}}$ can be considered as a `gain' term and is dependent on \ac{RIS} incidence and reflection angles, expressed as spherical coordinates $(\theta_\mathrm{i},\phi_\mathrm{i}, \theta_\mathrm{r},\phi_\mathrm{r})$. In this paper, we modified the formula (26) from \cite{bOkan} to express it as:
\begin{equation}
    \begin{split}
    PL^{\text{FS}}_{\text{RIS}}(\theta_\mathrm{i},\phi_\mathrm{i}, \theta_\mathrm{r},\phi_\mathrm{r}) = 
    \frac{\mathrm{G}_{\mathrm{Tx}} \mathrm{G}_{\mathrm{Rx}} \lambda^2  \sigma_{\mathrm{RIS}}
    }{(4\pi)^3 ( d_{\text{BS-RIS}} d_{\text{RIS-UE}} )^2},
    \end{split}
\end{equation}
\noindent
where $G_{\text{Tx}}$ is the gain of the transmitter antenna, $G_{\text{Rx}}$ is the gain of the receiver antenna, and $\sigma_{\text{RIS}}$ is the \ac{RCS} of \ac{RIS}, that encapsulates both beamsteering amplitude gains and structural amplitude losses of \ac{RIS} for a defined $(\theta_\mathrm{i}, \phi_\mathrm{i}, \theta_\mathrm{r},\phi_\mathrm{r} )$ combination, and expressed as:
\begin{align}
&\sigma_{\mathrm{RIS}}(\theta_i,\phi_i,\theta_r,\phi_r)
\approx\
\frac{4\pi\,A_{\mathrm{RIS}}^{2}}{\chi_\text{RIS}\lambda^{2}}\,
\cos^{2}\theta_i\,\cos^{2}\theta_r\; \nonumber \\&
\cdot\text{sinc}^{2}\!\left(\frac{\pi X}{\lambda}\left(\sin\theta_r\cos\phi_r+\sin\theta_i\cos\phi_i\right)\right)\, \nonumber \\&
\cdot\text{sinc}^{2}\!\left(\frac{\pi Y}{\lambda}\left(\sin\theta_r\sin\phi_r+\sin\theta_i\sin\phi_i\right)\right),
\label{eq:RCS}
\end{align}
\noindent
where $A_\mathrm{RIS}=X \times Y$ is the area of \ac{RIS} in $\text{m}^2$, and $\chi_{\text{RIS}}$ is the design-specific reflection loss per surface area term that differentiates  $\sigma_{\text{RIS}}$ from \ac{RCS} of a flat metal plate. Also, $\text{sinc}^2$ functions account for tangential phase-mismatch components between $(\theta_\mathrm{i},\phi_\mathrm{i})$ and $(\theta_\mathrm{r},\phi_\mathrm{r})$ along the \acp{RIS} X and Y axes.\par 
The $\sigma_{\text{RIS}}$ is typically measured under the free space conditions in an anechoic chamber for several common \ac{RIS} configurations and does not vary with distance. Measurements shown in Ch. 6.4 of \cite{bETSI} confirm that \ac{RIS} behaves approximately like a specular reflector, where \ac{RCS} is strongest near normal incidence and decays with the increasing incidence and reflection angles \cite{bSravan2}. This observation allows estimation of \ac{RCS} at all angles from the measured specular \ac{RCS} value by applying (\ref{eq:RCS}), allowing a scalable modeling of RIS in raytracing scenarios, based on a prior anechoic chamber measurement of its specular \ac{RCS}. However, this observation also highlights that the use of \acp{RIS} is not practical for angles exceeding $\pm$65$^\circ$.
\par



\begin{figure}[!t]
\centering

    \begin{turn}{0}
    \includegraphics[width=\columnwidth]{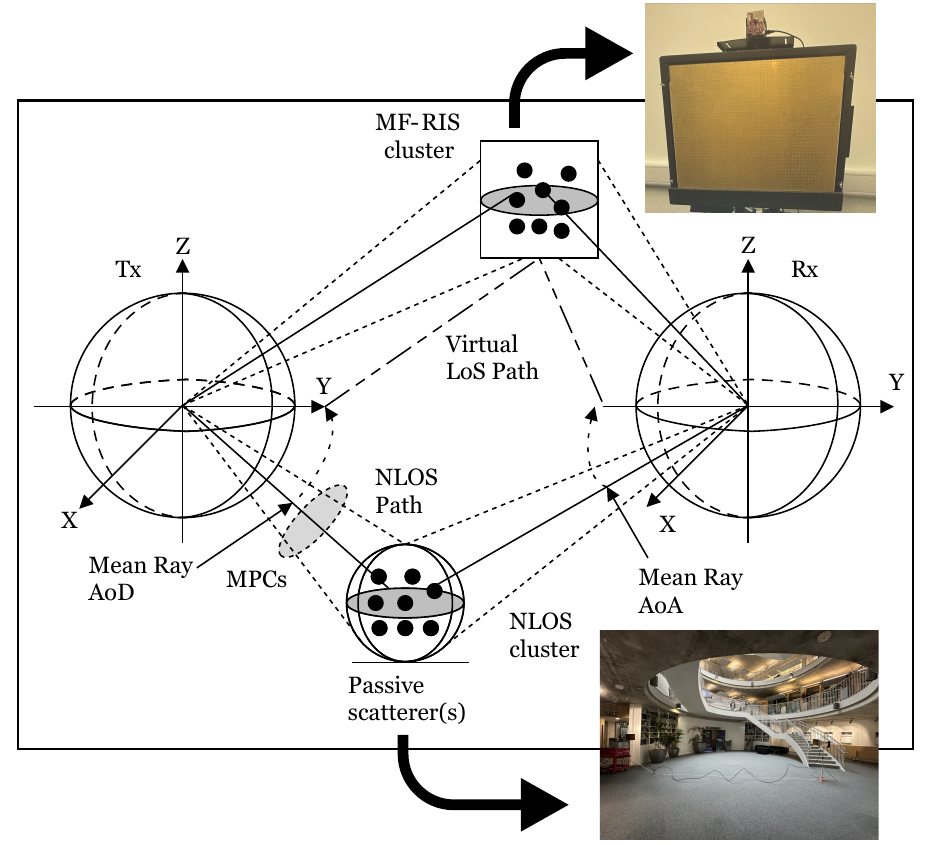} 
    \end{turn}

\caption{Overview of the \ac{GBSM} applied to model the fast-fading channel ($h_{\text{BS-RIS-UE}}$) in this work.}
\label{fig:GBSM}
\end{figure}

\subsection {Fast-Fading Channel Model}

To accurately model fast-fading of the \ac{MF-RIS}-assisted communication, we applied a \ac{GBSM} model with multiple levels of randomness, which is illustrated in Fig. \ref{fig:GBSM}. This model fragments the indoor environment into $Q$-number of clusters, where all moving objects are typically assumed to be humans. Non-moving clusters are the static objects in the environment, such as walls, doors, or office furniture. A \ac{MF-RIS} is also considered to be a static cluster, however, its response is reconfigured in several pre-defined ways according to its codebook, which is updated dynamically based on the moving cluster (assumed to be a UE) position. The resulting \ac{CIR} is defined in terms of its large-scale (shadow fading, Ricean K-factor, delay spread, angular spreads) and small-scale (individual path delays, cluster powers, angles of arrival and departure, cross-polarization power ratios, per cluster shadowing) parameters. This model also considers a sustained linear motion, which is captured by assigning a mean Doppler shift to each moving cluster, where each cluster consists of $I$-number of rays. In idealized conditions with perfect reflections, each cluster would consist of exactly one ray, however, in practice, this number is a variable. If the scatterer represented by the cluster is rigid, the relative motion of the rays arises purely from the geometry and the motion of the cluster itself, and can be directly computed from the geometric model and known movement parameters. Additionally, there may be internal motion within a cluster, such as the motion of limbs, introducing further relative movement among rays. This intra-cluster motion can be modeled using a Doppler spectrum assigned to the cluster, however, ray-tracing estimation of intra-cluster parameters is known to be inaccurate and should be obtained from the measurement data instead \cite{bMaltsev}. \par

We begin our analysis by expressing the high-level \ac{CIR} of the BS-RIS-UE path in $(\tau,t)$-domain:
\begin{align}
    &h_{u,s,q,i}^{\text{BS-RIS-UE}}(\tau,t) = \\& \nonumber
    \left(h_{\text{BS-RIS}}\circledast  h_{\text{RIS-UE}}\right)(\tau,t) + h_{\text{BS-UE}}(\tau,t),
    \label{eq:BS-RIS-UE}
\end{align}
where $\circledast$ indicates convolution.

Assuming the plane wave incidence and that both the \ac{BS} and the \ac{RIS} are static, the \ac{CIR} for each \ac{LoS} path between the BS terminal $s$ and the \ac{RIS} can be adopted directly from the \ac{3GPP} \ac{TR} 38.901 release 18.0.0 (7.5-29) \cite{3GPP1}:
\begin{align} \nonumber
    &h^{\text{BS-RIS}}_{s,1}(\tau,t) = \cdot\begin{bmatrix}
F_{\text{RIS},\theta}(\theta_{\text{ZoA}}, \phi_{\text{AoA}}) \\
F_{\text{RIS},\phi}(\theta_{\text{ZoA}}, \phi_{\text{AoA}})
\end{bmatrix}^{T} 
\begin{bmatrix}
1 & 0 \\
0 & -1
\end{bmatrix} \\[0.25em]& \nonumber
\cdot
\begin{bmatrix}
F_{\text{BS},s,\theta}(\theta_{\text{ZoD}}, \phi_{\text{AoD}}) \\
F_{\text{BS},s,\phi}(\theta_{\text{ZoD}}, \phi_{\text{AoD}})
\end{bmatrix} \exp\left(-j 2\pi \frac{ d_{\text{BS-RIS}}}{\lambda_0}\right)  \\[0.25em]& 
\cdot \exp\!\left(j 2\pi \frac{\hat{r}_{\text{BS},\text{LoS}}^{T} \vec{r}_{\text{BS},s}}{\lambda_0}\right)
\delta\left(\tau - \tau_1^{\text{BS-RIS}}\right),
\end{align}
where $F_{\text{RIS},\theta}$ and $F_{\text{RIS},\phi}$ are the re-radiation patterns of the \ac{RIS} in the direction of the spherical coordinates in a far-field $(\theta, \phi)$,  $F_{\text{BS},s,\theta}$ and $F_{\text{BS},s,\phi}$ are the equivalent radiation patterns of the transmit antenna $s$ in the direction of the spherical basis vectors $\theta$ and $\phi$ respectively, that both depend on the \ac{AoA}, the \ac{AoD}, the \ac{ZoA}, and the \ac{ZoD}, $\hat{r}_{\text{BS}}^T $ is the spherical unit vector with azimuth departure angle for each BS antenna element, $\lambda_0$ is the wavelength of a carrier signal, and $\delta(\cdot)$ is the Dirac's delta function. 
The measurement process of re-radiation patterns for the \ac{RIS} used in this paper and its results are described in Section 6.4 of \cite{bETSI}. It is important to note that \ac{RIS} re-radiation patterns change depending on the reflection angle, and should be either measured or simulated for each combination.

To express the \ac{CIR} of the RIS-UE channel, we derived the double-bounce \ac{3GPP}-compatible \ac{SV} model \cite{bSV}, which includes both the deterministic \ac{LoS} component and the \ac{NLoS} \acp{MPC} created by the \ac{RIS} in the environment:
\begin{align} \nonumber
&h^{\text{RIS-UE}}_{1,u}(\tau,t) = \sqrt{\frac{K_R}{K_R+1}}  
\begin{bmatrix}
F_{\text{UE},u,\theta}(\theta_{\text{ZoA}}, \phi_{\text{AoA}}) \\
F_{\text{UE},u,\phi}(\theta_{\text{ZoA}}, \phi_{\text{AoA}})
\end{bmatrix}^{T} \\[0.25em]& \nonumber
\cdot\begin{bmatrix}
1 & 0 \\
0 & -1
\end{bmatrix} 
\begin{bmatrix}
F_{\text{RIS},\theta}(\theta_{\text{ZoD}}, \phi_{\text{AoD}}) \\
F_{\text{RIS},\phi}(\theta_{\text{ZoD}}, \phi_{\text{AoD}})
\end{bmatrix} \exp\left(-j 2\pi \frac{ d_{\text{RIS-UE}}}{\lambda_0}\right) \\[0.25em]& 
\cdot \exp\left( j 2\pi \frac{\hat{r}_{\text{UE},\text{LoS}}^T \vec{r}_{\text{UE},u}}{\lambda_0} \right) \exp\left( j 2\pi \frac{\hat{r}_{\text{UE},\text{LoS}}^T \vec{v}}{\lambda_0} t \right) \nonumber \\[0.25em]& 
\cdot
\delta\left(\tau - \tau_1^{\text{RIS-UE}}\right) \nonumber\\[0.25em]& 
 + \sqrt{\frac{1}{K_R+1}}
\sum_{q=1}^Q\sum_{i=1}^{I_q}\sqrt{\frac{P_q}{I_q}}
\begin{bmatrix}
F_{\text{UE},u,\theta}(\theta_{q,i,\text{ZoA}}, \phi_{q,i,\text{AoA}}) \\
F_{\text{UE},u,\phi}(\theta_{q,i,\text{ZoA}}, \phi_{q,i,\text{AoA}})
\end{bmatrix}^{T} \nonumber \\[0.25em]
&\cdot
\begin{bmatrix}
\exp(j \Phi_{q,i}^{\theta\theta}) & \sqrt{\kappa_{q,i}^{-1}} \exp(j \Phi_{q,i}^{\theta\phi}) \\
\sqrt{\kappa_{q,i}^{-1}} \exp(j \Phi_{q,i}^{\phi\theta}) & \exp(j \Phi_{q,i}^{\phi\phi})
\end{bmatrix} 
\nonumber \\[0.25em]& 
\cdot
\begin{bmatrix}
F_{\text{RIS},\theta}(\theta_{\text{q,i,ZoD}}, \phi_{\text{q,i,AoD}}) \\
F_{\text{RIS},\phi}(\theta_{\text{q,i,ZoD}}, \phi_{\text{q,i,AoD}})
\end{bmatrix} 
\exp\left( j 2\pi \frac{\hat{r}_{\text{UE},q,i}^T \vec{r}_{\text{UE},u}}{\lambda_0} \right) \nonumber \\[0.25em]& \cdot \exp\left( j 2\pi \frac{\hat{r}_{\text{UE},\text{q,i}}^T \vec{v}}{\lambda_0} t \right) 
\delta\left(\tau - \tau_{q}^{\text{RIS-UE}} -\tau_{q,i}^{\text{RIS-UE}}\right),
\label{eq:ris-ue}
\end{align}
\noindent
where $K_R$ is the Ricean K-factor in $\text{dB}$ which refers to a power ratio between the deterministic component and the stochastic components (i.e. the main reflected path and other \acp{MPC}), $Q$ is the number of clusters, $I_q$ is the number of rays per cluster, $\hat{r}_{\text{UE}}^T $ is the spherical unit vector with azimuth departure angle for each \ac{UE} antenna element, $\phi_{\text{AoD}}$ and elevation departure angle $\theta_{\text{ZoD}}$, $\vec{r}_{\text{UE},u}$ is the location vector of receive antenna at the \ac{UE}, $\vec{v}$ is the velocity vector of the \ac{UE} with speed $v$ and travel direction angles $\left(\phi_v,\theta_v\right)$, $\Phi_{q,i}^{\theta\phi}$ denote random initial phases for path $i$ in cluster $q$ for four different polarization combinations in  \acp{CIR}, and $k$ is the \ac{XPR} for each ray $i$ of each cluster $q$, such that:
\begin{equation}
    k_{q,i} = 10^{X_{q,i}/10},
\end{equation}
where $X_{q,i} \sim N \left(\mu_{\text{XPR}},\sigma_{\text{XPR}}\right)$ is Gaussian (i.e. normal) distributed with $\mu_{\text{XPR}}$, and $\sigma_{\text{XPR}}$ values given in Table 7.5-6 of \cite{3GPP1}, and $P_q$ are the cluster powers defined as:
\begin{equation}
    P_q= \frac{P^\prime_q}{\sum^Q_{q=1}P^\prime_q},
\end{equation}
\begin{equation}
    P^\prime_q = \exp{\left( -\tau_q \frac{\omega_\tau-1}{\omega_\tau DS} \right)\cdot10^{\frac{-Z_n}{10}}},
\end{equation}
where $DS$ is the delay spread, $\omega_\tau$ is the decay rate, and $Z_{n} \sim N (0,\zeta^2)$ is the per-cluster shadowing term in $\text{dB}$, which depends on the delay distribution. \par

\begin{figure*}[!t]
\begin{align}
&h_{u,s}^{\text{BS-RIS-UE}}(\tau,t)
= \sqrt{\frac{K_R}{K_R+1}}\,
\begin{bmatrix}
F_{\text{UE},u,\theta}(\theta_{\text{ZoA}}, \phi_{\text{AoA}}) \\
F_{\text{UE},u,\phi}(\theta_{\text{ZoA}}, \phi_{\text{AoA}})
\end{bmatrix}^{T}
\begin{bmatrix}
F_{\text{RIS},\theta}\!\left(\theta_{\text{ZoA}}, \phi_{\text{AoA}},\theta_{\text{ZoD}}, \phi_{\text{AoD}}\right) &
F_{\text{RIS},\phi}\!\left(\theta_{\text{ZoD}}, \phi_{\text{AoD}},\theta_{\text{ZoA}}, \phi_{\text{AoA}} \right) \\
F_{\text{RIS},\phi}\!\left(\theta_{\text{ZoA}}, \phi_{\text{AoA}},\theta_{\text{ZoD}}, \phi_{\text{AoD}}\right) &
F_{\text{RIS},\theta}\!\left(\theta_{\text{ZoD}}, \phi_{\text{AoD}},\theta_{\text{ZoA}}, \phi_{\text{AoA}} \right)
\end{bmatrix} \notag \\[0.25em]
& 
\cdot 
\begin{bmatrix}
F_{\text{BS},s,\theta}(\theta_{\text{ZoD}}, \phi_{\text{AoD}}) \\
F_{\text{BS},s,\phi}(\theta_{\text{ZoD}}, \phi_{\text{AoD}})
\end{bmatrix}
\exp\!\left(-j 2\pi \frac{d_{\text{BS-RIS}} + d_{\text{RIS-UE}}}{\lambda_0}\right)
\exp\!\left(j 2\pi \frac{\hat{r}_{\text{UE},\text{vLoS}}^{T} \vec{r}_{\text{UE},u}}{\lambda_0}\right)
\exp\!\left(j 2\pi \frac{\hat{r}_{\text{UE},\text{vLoS}}^{T} \vec{v}}{\lambda_0} t\right)  \exp\!\left(j 2\pi \frac{\hat{r}_{\text{BS},\text{vLoS}}^{T} \vec{r}_{\text{BS},s}}{\lambda_0}\right) \notag \\[0.25em]&
\cdot \delta\!\left(\tau - \tau_1^{\text{BS-RIS-UE}}\right)  + \sqrt{\frac{1}{K_R+1}}\,
\Bigg[
\sum_{q=1}^{Q}\sum_{i=1}^{I_q} \sqrt{\frac{P_q}{I_q}}\,
\begin{bmatrix}
F_{\text{UE},u,\theta}(\theta_{q,i,\text{ZoA}}, \phi_{q,i,\text{AoA}}) \\
F_{\text{UE},u,\phi}(\theta_{q,i,\text{ZoA}}, \phi_{q,i,\text{AoA}})
\end{bmatrix}^{T} 
\begin{bmatrix}
\exp(j \Phi_{q,i}^{\theta\theta}) & \sqrt{\kappa_{q,i}^{-1}} \exp(j \Phi_{q,i}^{\theta\phi}) \\
\sqrt{\kappa_{q,i}^{-1}} \exp(j \Phi_{q,i}^{\phi\theta}) & \exp(j \Phi_{q,i}^{\phi\phi})
\end{bmatrix} \notag \\[0.25em]&\cdot 
\begin{bmatrix}
F_{\text{RIS},\theta}(\theta_{\text{q,i,ZoD}}, \phi_{\text{q,i,AoD}}) \\
F_{\text{RIS},\phi}(\theta_{\text{q,i,ZoD}}, \phi_{\text{q,i,AoD}})
\end{bmatrix} 
\exp\!\left(j 2\pi \frac{\hat{r}_{\text{UE},q,i}^{T} \vec{r}_{\text{UE},u}}{\lambda_0}\right)
\exp\!\left(j 2\pi \frac{\hat{r}_{\text{UE},\text{q,i}}^{T} \vec{v}}{\lambda_0} t\right) 
\delta\!\left(\tau - \tau_{q}^{\text{BS-RIS-UE}} - \tau_{q,i}^{\text{BS-RIS-UE}}\right) \notag \\[0.25em]
&
 + \sum_{o=1}^{O} \sum_{w=1}^{W_o} \sqrt{\frac{P_o}{W_o}}\,
\begin{bmatrix}
F_{\text{UE},u,\theta}(\theta_{o,w,\text{ZoA}}, \phi_{o,w,\text{AoA}}) \\
F_{\text{UE},u,\phi}(\theta_{o,w,\text{ZoA}}, \phi_{o,w,\text{AoA}})
\end{bmatrix}^{T}
\begin{bmatrix}
\exp(j \Phi_{o,w}^{\theta\theta}) & \sqrt{\kappa_{o,w}^{-1}} \exp(j \Phi_{o,w}^{\theta\phi}) \\
\sqrt{\kappa_{o,w}^{-1}} \exp(j \Phi_{o,w}^{\phi\theta}) & \exp(j \Phi_{o,w}^{\phi\phi})
\end{bmatrix}
\begin{bmatrix}
F_{\text{BS},s,\theta}(\theta_{o,w,\text{ZoD}}, \phi_{o,w,\text{AoD}}) \\
F_{\text{BS},s,\phi}(\theta_{o,w,\text{ZoD}}, \phi_{o,w,\text{AoD}})
\end{bmatrix}
\notag \\[0.25em]
& \cdot
\exp\!\left(j 2\pi \frac{\vec{r}_{\text{UE},o,w}^{T} \vec{r}_{\text{UE},u}}{\lambda_0}\right)
\exp\!\left(j 2\pi \frac{\vec{r}_{\text{UE},o,w}^{T} \vec{v}}{\lambda_0} t\right) \exp\!\left(j 2\pi \frac{\hat{r}_{\text{BS},\text{o,w}}^{T} \vec{r}_{\text{BS},s}}{\lambda_0}\right)
\delta\!\left(\tau - \tau_{o}^{\text{BS-UE}} - \tau_{o,w}^{\text{BS-UE}}\right)
\Bigg]. \label{eq:bs-ris-ue-full}
\end{align}
\hrulefill
\end{figure*}

\begin{figure*}[!t]

        \begin{subfigure}[b]{0.495\textwidth}
            \centering
            \includegraphics[width=\textwidth]{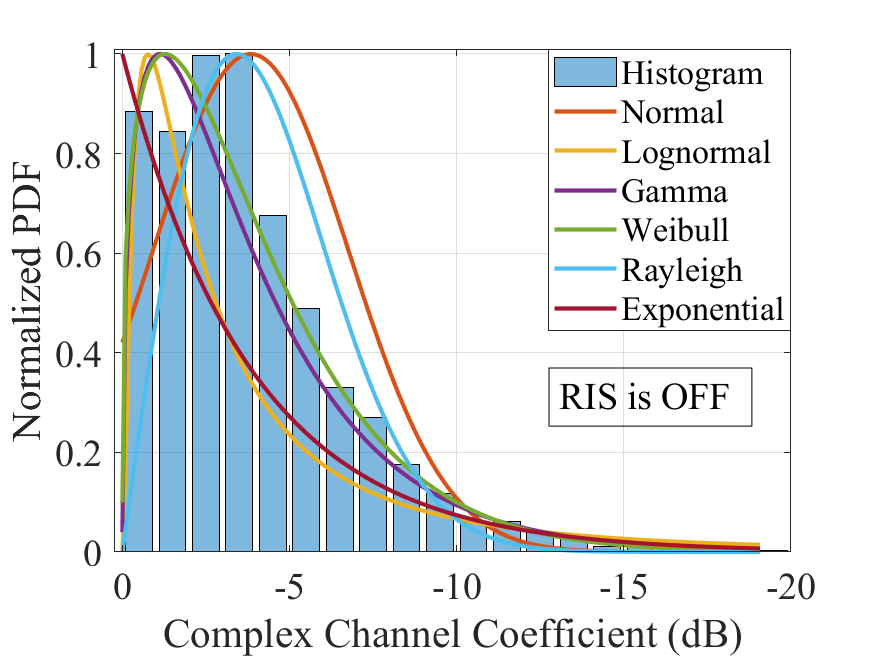} 
            \caption{}
        \end{subfigure}
        \begin{subfigure}[b]{0.495\textwidth}
            \centering
            \includegraphics[width=\textwidth]{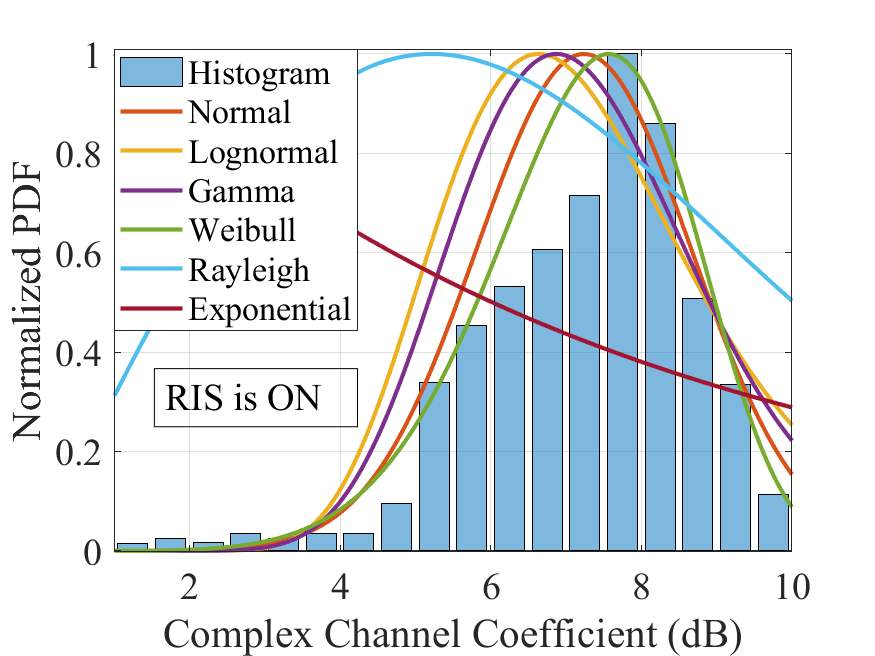} 
            \caption{}
        \end{subfigure}

\caption{Channel histograms for the UE coordinate (40$^\circ$, 90$^\circ$, 2m) plotted on measured \ac{PDP} with: a) \ac{RIS} switched OFF, showing fading in the BS-UE NLoS channel, and b) RIS switched ON, showing fading in the cascaded BS-RIS-UE channel i.e. the \ac{vLoS}.}
\label{fig:BS-RIS-UE}
\end{figure*}




Our measurements confirm that the effects of the \ac{NLoS} $h_{\text{BS-UE}}$ channel should not be ignored, as the \ac{NLoS} component from the \ac{BS} is inevitably present at the \ac{UE} due to \acp{MPC} and scattering in the environment with $O$-number of clusters and $W_o$-number of rays in these clusters, that may be different from the rays and clusters in the RIS-UE path. This is especially evident in the Sub-6~GHz indoor channel when $h_{\text{BS-RIS}}$ and $h_{\text{RIS-UE}}$ are both \ac{LoS} paths, thereby creating the \ac{vLoS} with the \ac{RIS}. Therefore, we also consider the \ac{CIR} of the \ac{NLoS} $h_{\text{BS-UE}}$ path, derived in \ac{3GPP} \ac{TR} 38 901 (7.5-22) \cite{3GPP1}. Then, equation (\ref{eq:ris-ue}) can be expanded to describe the \ac{CIR} of the cascaded $h_\text{BS-RIS-UE}$ channel as the finalized expression in (\ref{eq:bs-ris-ue-full}). Note that the basis-reconciliation sign flip $\left(\text{diag}\left(1,-1\right)\right)$ cancels out, as it is applied on both sides of the \ac{RIS} re-radiation patterns, and the \ac{RIS} is assumed to be a scalar (i.e., diagonal). Also note that this formulation assumes that the $h_\text{BS-RIS}$ channel is a \ac{LoS}, the $h_\text{BS-UE}$ channel is \ac{NLoS} (i.e. scattered), and neither the \ac{BS} nor the \ac{RIS} is moving. \par
The derived \ac{CIR} is compatible with the \ac{3GPP}-developed framework for \ac{ISAC}, and can be used in channel estimation and wireless ray-tracing simulation platforms, such as the ones described in \cite{b8, b9, b10}. The resulting channel can be visualized with a \ac{PDF} of the measured  $h_{\text{BS-RIS-UE}}$, as shown in Fig. \ref{fig:BS-RIS-UE} (b), with the \ac{UE} coordinate $(\theta,\phi,d) = (40^\circ, 90^\circ, 2\text{m})$, and compared with a no-\ac{RIS} scenario in Fig. \ref{fig:BS-RIS-UE} (a). The \ac{PDF} significantly differs from the conventional Rayleigh distribution with the \ac{RIS}. Instead, the channel distribution is most closely matched to the Weibull distribution in this case, but the log-normal distribution is the most fitting when the bandwidth is reduced to 200 MHz.

\section{\ac{MF-RIS}-enabled \ac{RCC} system}\label{ch4}

Through the combination of the analysis shown in Sections \ref{ch2} and \ref{ch3}, this section provides the analysis of the complex \ac{MF-RIS} system with decoupled \ac{RCC} tasks.  

\subsection{\ac{MF-RIS} Algorithm for Localization of a Moving User}

 \begin{figure}[!t]
\centering

    \begin{turn}{0}
    \includegraphics[width=0.9\columnwidth]{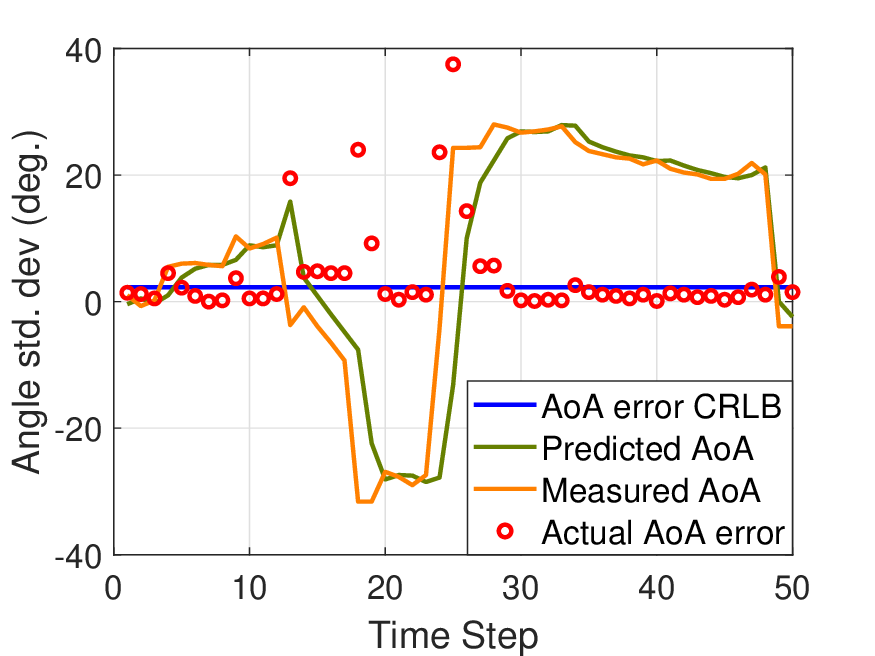} 
    \end{turn}

\caption{Horizontal \ac{AoA} Tracking: \ac{CRLB} vs predicted \ac{AoA} vs measured \ac{AoA} from a \ac{UE} moving in the radar's \ac{FoV}, and the resulting \ac{AoA} error.}
\label{fig:BCRLB}
\end{figure}

The Kalman filter prediction can be implemented on each separate term of the \ac{UE} position coordinate $\left(\theta^\prime_k, \phi^\prime_k,r^\prime_k\right)$ detected by the \ac{mmWave} radar. To illustrate, in Fig. \ref{fig:BCRLB}, the \ac{CRLB} of the theoretical minimum \ac{AoA} error is plotted for the \ac{mmWave} radar, previously derived in \cite{bCRLB}, and compared against the predicted Kalman filter output $(\theta^\prime_k)$, as well as the measured horizontal \ac{AoA} $(\theta_k)$, therefore showing the actual \ac{AoA} estimation error with red circles. This measurement was obtained with the user moving within the \ac{FoV} of the radar at a 2~m distance. The \ac{mmWave} radar hardware is based on the \Ac{TI} IWR6843ISK 60~GHz \ac{MIMO} radar development kit with 3 \ac{TX} antennas, 4 \ac{RX} antennas, transmit power $12~ \textrm{dBm}$, and $\mathrm{10~MHz}$ \ac{IF} bandwidth. As can be seen from the figure, a sudden movement or a trajectory change of the UE can introduce a significant \ac{AoA} error, however, the Kalman filter prediction error is greatly reduced within 3-4 frame durations, after which the \ac{AoA} error becomes aligned with the \ac{CRLB}, while the \ac{UE} is static or moving with a constant velocity. Note that the frame duration (i.e. prediction latency) depends on the \ac{SBC} speed and loading of the \ac{MCU}. The average prediction latency 127~ms was measured over one minute with the \ac{MF-RIS} ranging between 27~ms (fastest timestamp) and 297~ms (slowest timestamp) with the average horizontal \ac{AoA} $(\theta_k)$ error 6.47$^\circ$. This result is deemed appropriate for an indoor-centric setup. However, in an outdoor-centric setup such performance might be sub-optimal, therefore requiring more sophisticated hardware. The developed \ac{MF-RIS} phase update algorithm is summarized in Algorithm~\ref{alg:mmwave_ris}\footnotemark[1].

\footnotetext[1]{The developed \ac{MF-RIS} localization Python-based code and the design files are available on GitHub: https://github.com/DREMCLTD/FlexiDAS-RIS.}

\begin{algorithm}[htbp]
\DontPrintSemicolon
\SetAlgoLined
\SetKwInOut{Input}{Input}
\SetKwInOut{Output}{Output}

\Input{Radar sensor data stream, serial ports, configuration commands}
\Output{Estimated \ac{AoA} and \ac{MF-RIS} phase synthesis}

\textbf{Initialization:} Set constants, buffers, Kalman filter, and \ac{MF-RIS} image mappings\;
Open \ac{CLI} and data serial ports\;

\textbf{Sensor Setup:} \\
\ForEach{command in configCommands}{
    Send command via the \ac{CLI} serial port\;
    Wait 10 ms\;
}

\textbf{Parse Configuration:} \\
Extract radar parameters (e.g., range resolution, Doppler bins)\;

\While{application is running}{
    Read bytes from the data port into the buffer to find the start of each packet\;
    \If{magic word found}{
        Parse frame and extract: $x$, $y$, $z$, azimuth, \ac{SNR}, velocity\;
    }

    Add frame to rolling buffer; aggregate multiple frames\;
    Filter low-\ac{SNR} and background points\;

    \If{clustering enabled}{
        Apply \ac{DBSCAN} on filtered points\;
        \If{valid clusters found}{
            Select the best cluster (largest moving)\;
            \If{Kalman not initialized}{
                Initialize Kalman filter\;
            }
            \Else{
                Predict and update with centroid + \ac{AoA}\;
            }
            Draw a bounding box and annotate \ac{AoA}\;
            \If{\ac{AoA} stable for $N$ frames}{
                Map \ac{AoA} to image index\;
                Send index to \ac{MF-RIS} via serial port\;
                    \If{stored index is different}{
                    Perform \ac{MF-RIS} phase synthesis\;
                    }
                Reset confirmation counter\;
            }
        }
        \ElseIf{previous cluster exists}{
            Predict new position using last known velocity\;
            Update GUI accordingly\;
        }
    }
    \Else{
        Display raw filtered points without clustering\;
    }
}

\caption{\ac{mmWave} radar-based \ac{AoA} estimation and \ac{MF-RIS} phase update}
\label{alg:mmwave_ris}
\end{algorithm}

\subsection{\ac{MF-RIS}-Induced Channel Hardening Effect}

The instantaneous \ac{SNR} of the system in (\ref{eq:bs-ris-ue-full}) is derived as:
\begin{equation}
    \text{SNR} = \frac{P^{\text{Tx}}_{s}}{\eta^2}\left(\sum^L_{l=1} \left|h_{u,s}^{\text{BS-RIS-UE}}(\tau,t)\right|\right)^2,
    \label{eq:SNR}
\end{equation}
where $P^{\text{Tx}}_{s}$ is the transmit power, $\eta$ is the noise variance, and $L$ defines the number of channels. 
It follows from (\ref{eq:SNR}) that randomness in $h_{\text{BS-RIS}}$, $h_{\text{RIS-UE}}$, and $h_{\text{BS-UE}}$ causes \ac{SNR} fluctuations, degrading communication performance. When these variations average out, the channel exhibits hardening \cite{CH1, CH2, CH3}, meaning the random \ac{SNR} can be approximated by a deterministic term. Practically, this implies $\text{SNR} \approx L^2$ times a constant for large $L$, where $\left(h_{\text{BS-RIS}},h_{\text{RIS-UE}},h_{\text{BS-UE}}:l=1,...,L\right)$ are random sequences converging to deterministic values. Therefore, channel hardening occurs when $h_{\text{BS-RIS-UE}}$ dominates over $h_{\text{BS-UE}}$.

Just as random channel gains cause \ac{SNR} fluctuations, random multipath delays cause fluctuations in the instantaneous coherence bandwidth $B_c(t)$, as outlined in Chapter 4.5 of \cite{Rappaport} with several practical examples. To summarize, in \ac{NLoS} with many reflections, the \ac{RMS} delay spread $\tau_{\text{RMS}}$ is large, giving a small $B_c(t)$ and strong frequency selectivity. However, in \ac{RIS}-enabled \ac{vLoS} with small delay spread, $B_c$ is large instead, giving near flat fading. With mobility and path geometry changes, delay spread alters and thus $B_c(t) \approx 1/5 \times \tau_{\text{RMS}}(t)$, which is an estimate that assumes that the frequency correlation between subcarriers is equal to 0.5 (i.e., moderately similar channel gains/phases). Hence, just as (\ref{eq:SNR}) shows power fluctuations from random gains, $B_c(t)$ captures frequency-domain variability, explaining unstable throughput in mobile \ac{NLoS} \ac{5G} \ac{NR} channels.

To reduce the frequency selectivity of the channel, we propose to minimize the delay spread $\tau_{\text{RMS}}$ by maximizing $B_c$ through real-time \ac{MF-RIS} system configuration. We can formulate the corresponding optimization problem in terms of the set of phase shifts $\left(\Psi_{m,n}\right)$ on \ac{MF-RIS}. Similar to sum-rate maximization, the problem can be written as an arg-minimization of the RMS delay spread:
\begin{align}
&
\mathbf{\Psi_{m,n}^{\star}} = \underset{{\Psi_{m,n}}}{\arg\min} \ \tau_{\text{RMS}}(t),  \\& \nonumber
\text{s.t.} \ \Psi_{m,n} \in [0^\circ, 360^\circ], \quad \forall m,n.
\end{align}

Equivalently, this optimization problem can be written as the maximization of the coherence bandwidth:
\begin{align}
\mathbf{\Psi_{m,n}^{\star}} = \underset{{\Psi_{m,n}}}{\arg\max}\ B_c(t) = \underset{{\Psi_{m,n}}}{\arg\max} \ \frac{1}{5~\tau_{\text{RMS}}}{(t)}~,
\end{align}
In other words, by dynamically adjusting phase shifts on the \ac{MF-RIS}, the channel’s delay spread is compressed and the channel appears flat over a larger bandwidth for a given \ac{UE}, thus preventing throughput fluctuations when it is moving.\par

From this analysis, it can be observed that the optimization of both \ac{SNR} and the coherence bandwidth for \ac{MF-RIS}-assisted communication can cause a slight trade-off under specific circumstances. For example, if there are multiple paths (e.g., a direct \ac{LoS} path and a \ac{MF-RIS}-created \ac{vLoS} path with a longer delay), maximizing \ac{SNR} might involve constructively adding both – but that introduces a larger delay gap between them. Minimizing delay spread might instead favor one dominant path and suppress secondary paths. Then, the codebook design might sacrifice some power from a late-arriving path to keep the channel more time-concentrated. Therefore, there is a balance between power maximization and delay alignment. In this work, we focus on the latter and use the standard deviation of channel throughput as the performance metric for evaluating the \ac{MF-RIS}-induced channel hardening effect. 

\begin{figure}[!t]
\centering
    \begin{turn}{0}
    \includegraphics[clip, trim=6cm 1cm 6cm 0cm,width=\columnwidth]{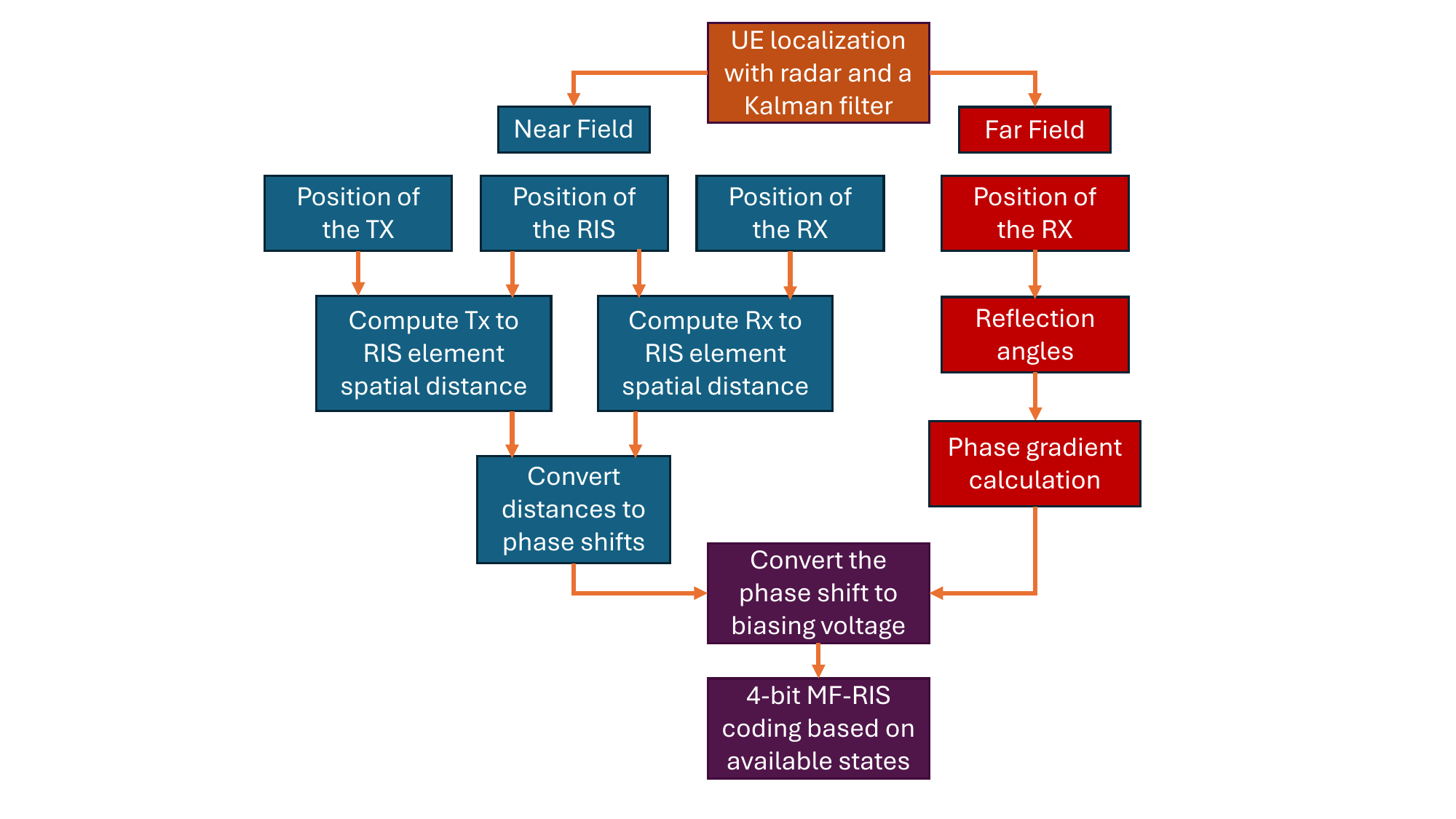} 
    \end{turn}
\caption{\ac{MF-RIS} phase synthesis algorithm for near-field and far-field domains.}

\label{fig:alg}
\end{figure}

\begin{figure*}[!t]
\centering

    \begin{minipage}{1\textwidth} 
        \centering
        \begin{subfigure}[b]{0.32\textwidth}
            \centering
            \includegraphics[width=\textwidth]{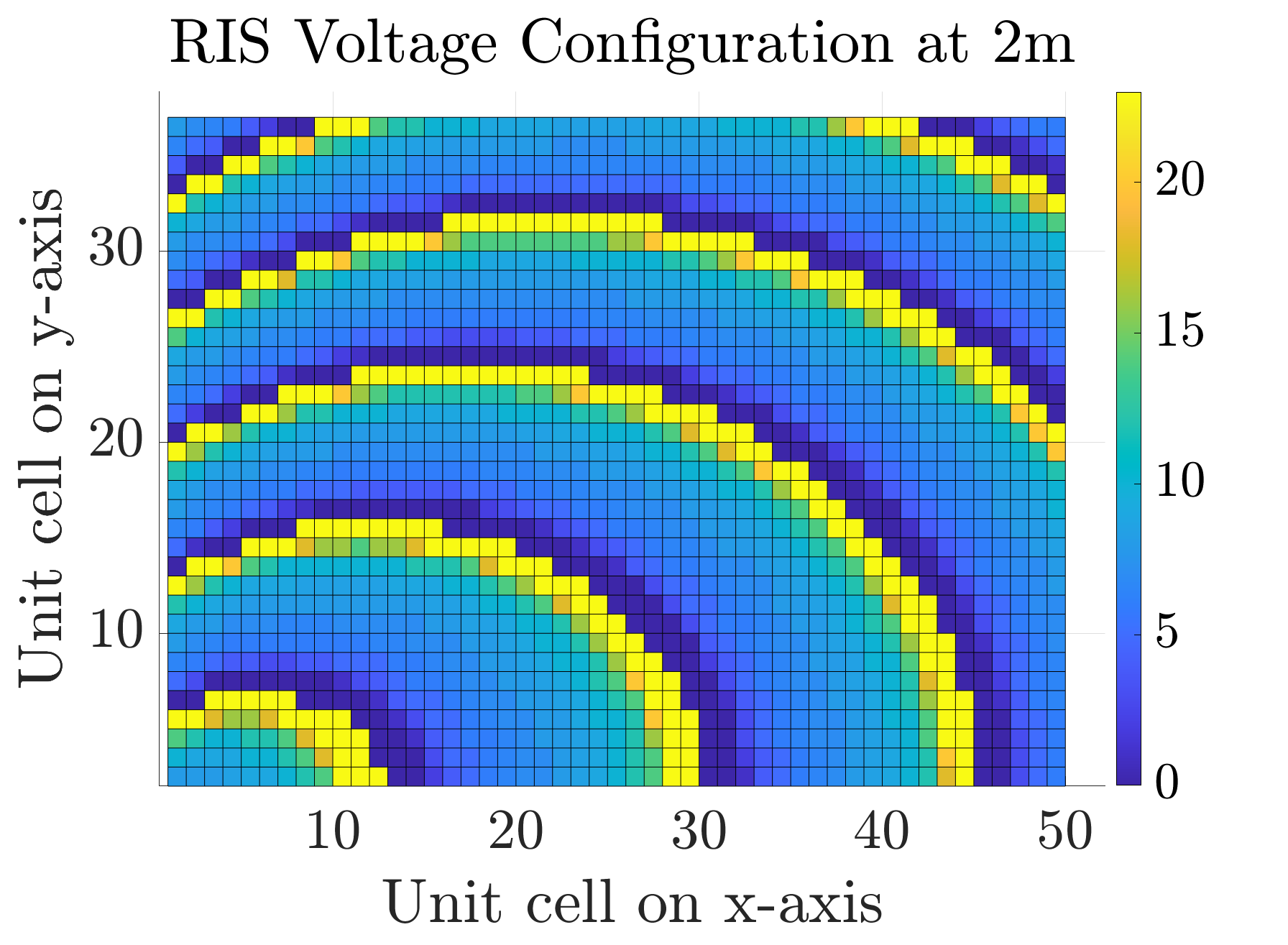} 
            \caption{}
        \end{subfigure}
        \hfill
        \begin{subfigure}[b]{0.32\textwidth}
            \centering
            \includegraphics[width=\textwidth]{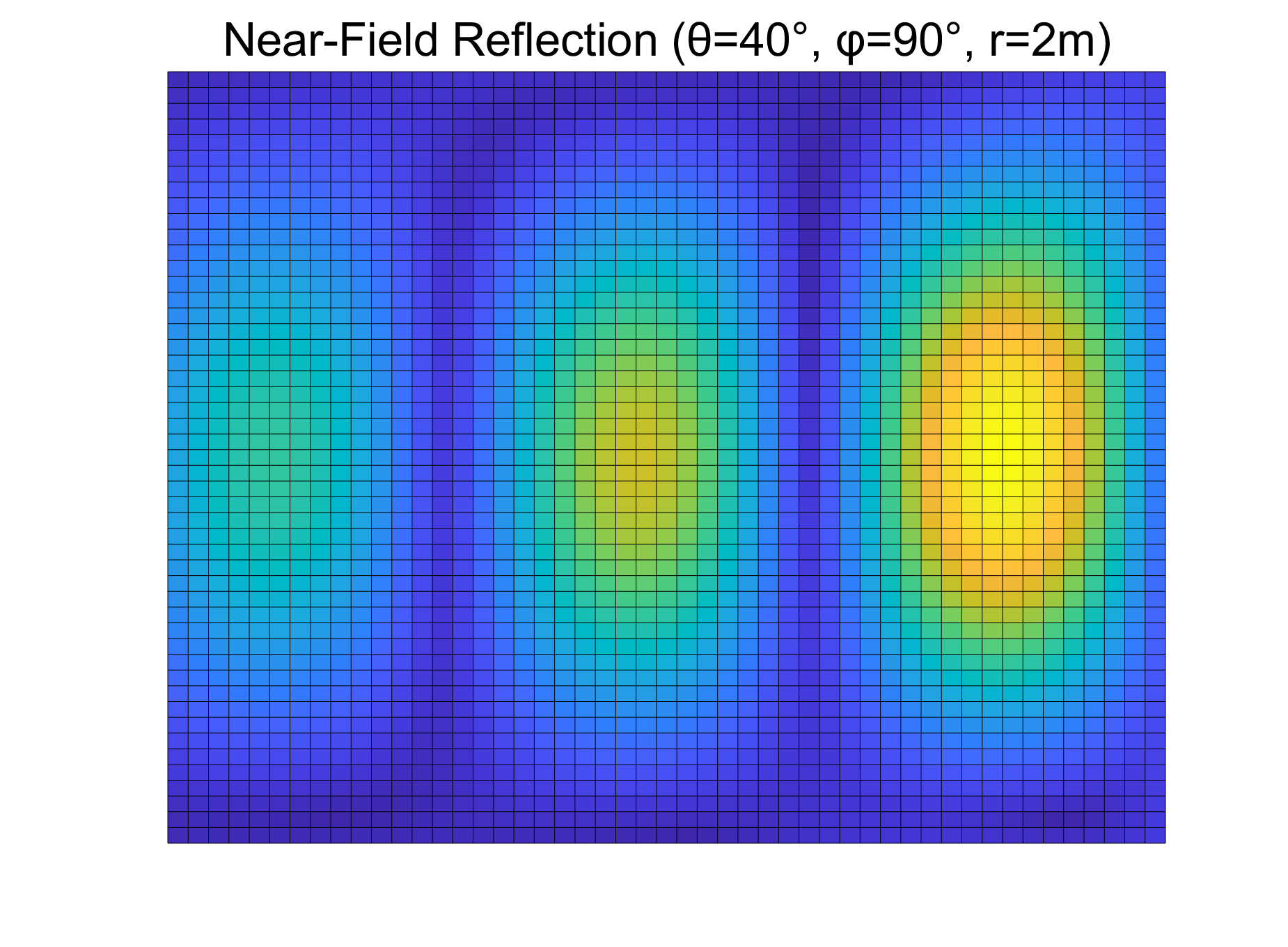} 
            \caption{}
        \end{subfigure}
        \hfill
        \begin{subfigure}[b]{0.32\textwidth}
            \centering
            \includegraphics[width=\textwidth]{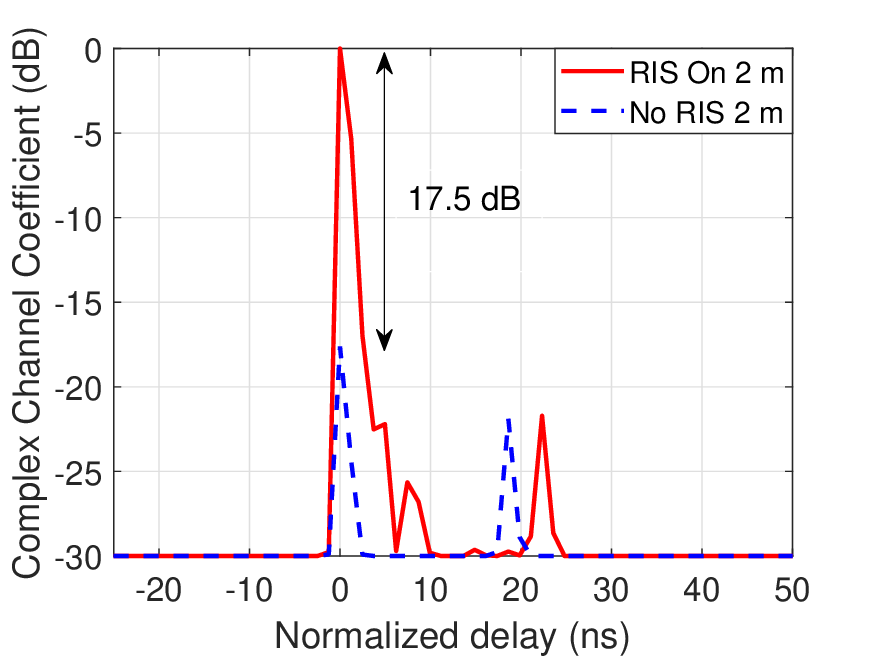} 
            \caption{}
        \end{subfigure}
    \end{minipage}

\caption{\ac{RIS} near-field phase synthesis for the reflection angle $\theta=40^\circ$, $\phi=90^\circ$ at 2~m distance ($d_{\text{RIS-UE}}$): a) varactor diode voltage, b) simulated beam pattern, c) the measured peak of the \ac{PDP} in the near-field for both \ac{RIS} on and no-\ac{RIS} cases.}
\label{near}
\end{figure*}

\begin{figure*}[!t]
\centering

    \begin{minipage}{1\textwidth} 
        \centering
        \begin{subfigure}[b]{0.32\textwidth}
            \centering
            \includegraphics[width=\textwidth]{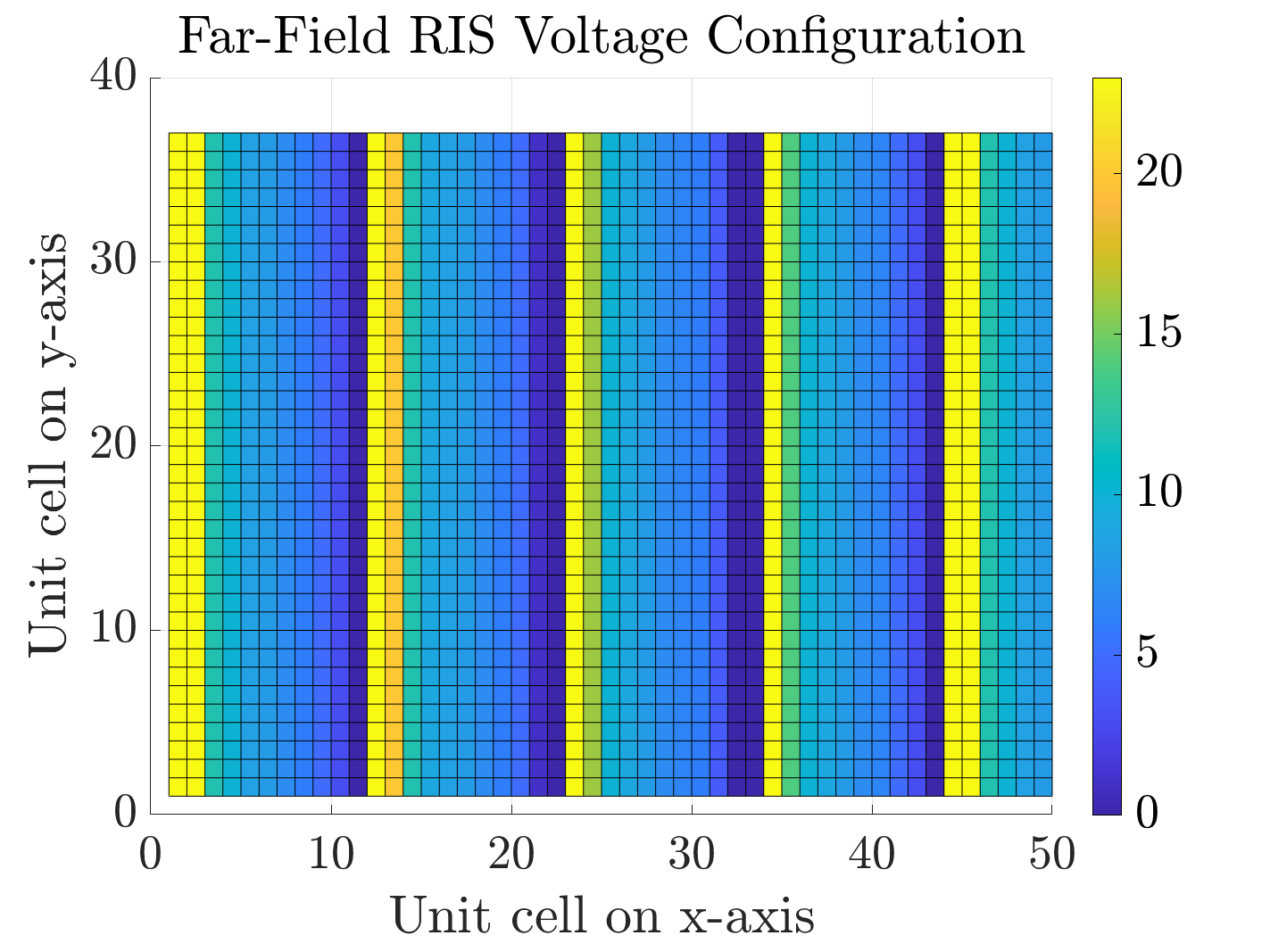} 
            \caption{}
        \end{subfigure}
        \hfill
        \begin{subfigure}[b]{0.32\textwidth}
            \centering
            \includegraphics[width=\textwidth]{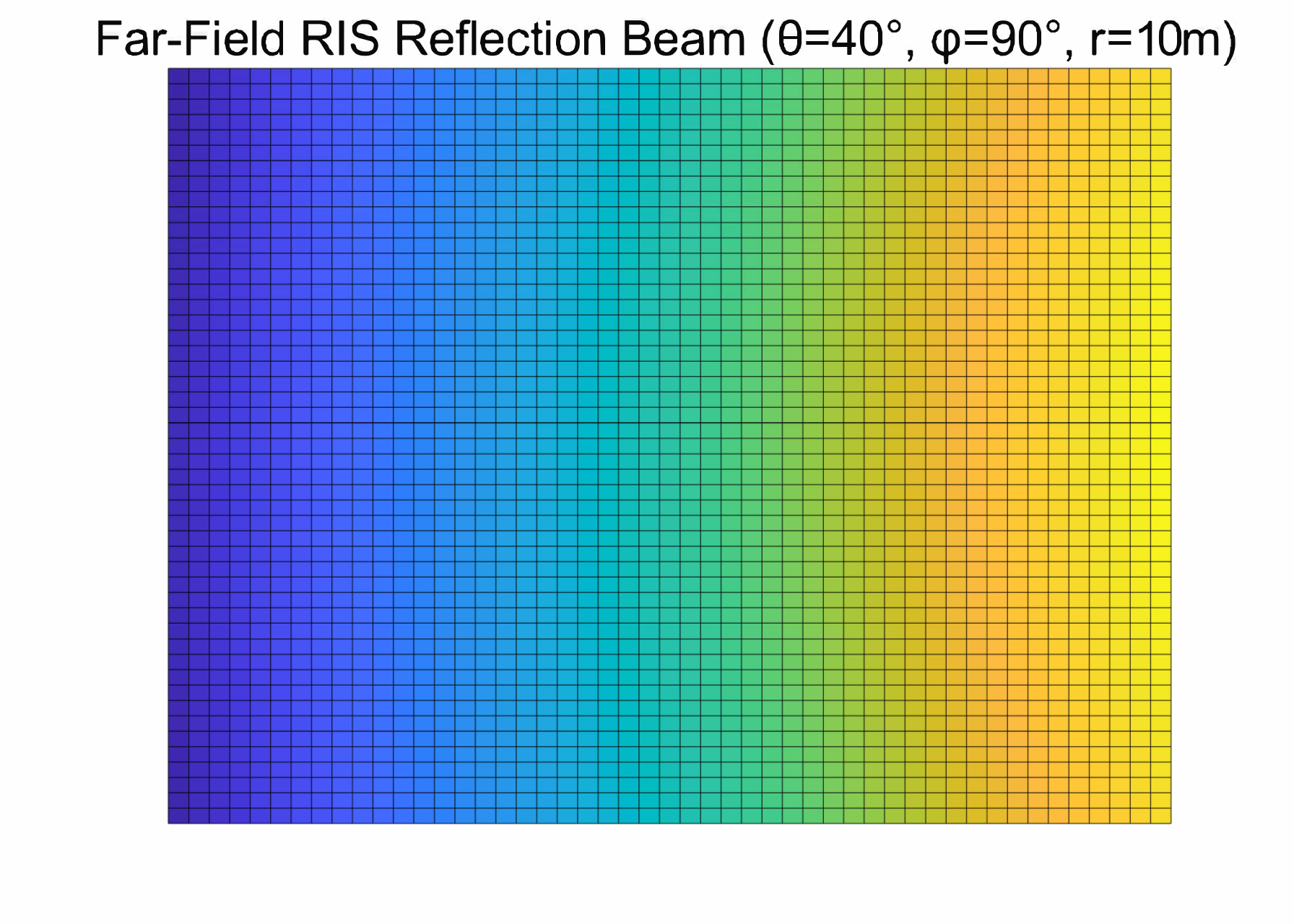} 
            \caption{}
        \end{subfigure}
        \hfill
        \begin{subfigure}[b]{0.32\textwidth}
            \centering
            \includegraphics[width=\textwidth]{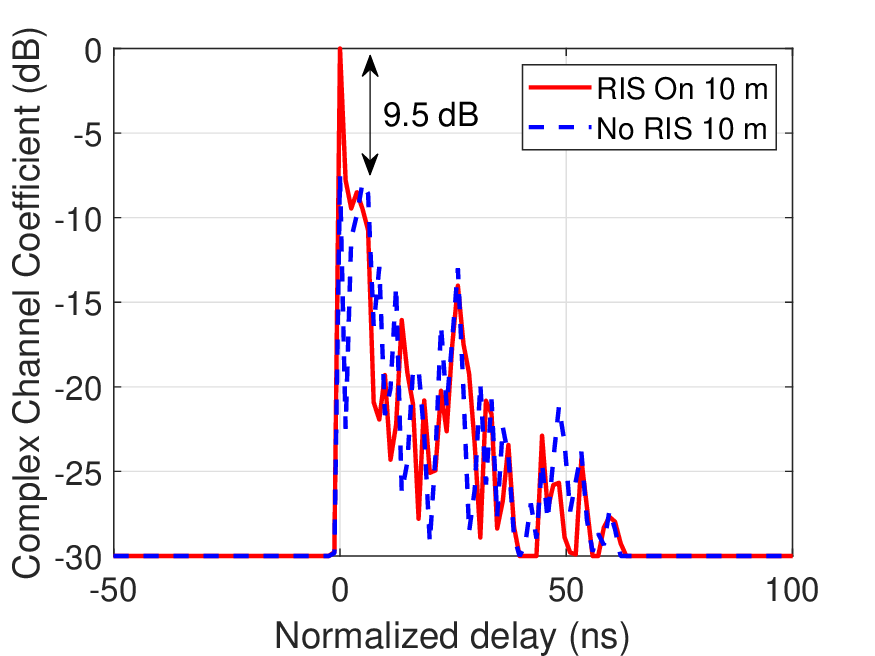} 
            \caption{}
        \end{subfigure}
    \end{minipage}

\caption{\ac{RIS} far-field phase synthesis for the reflection angle $\theta=40^\circ$, $\phi=90^\circ$: a) varactor diode voltage, b) simulated beam pattern, c) the measured peak of the \ac{PDP} in the far-field at 10 m distance ($d_{\text{RIS-UE}}$) for both \ac{RIS} on and no-\ac{RIS} cases.}
\label{far}
\end{figure*}

\section{\ac{MF-RIS} Phase Synthesis for Near-Field and Far-Field Domains}\label{ch5}

The propagation characteristics of the \ac{RIS}-assisted channel depend on EM interactions between the incident, reflected, and scattered waves. In an idealized far-field scenario, the \ac{RIS} reflection follows a planar wavefront-based uniform phase gradient model, while near-field scenarios require more complex modeling. Specifically, the near-field focusing effect is influenced by spherical wavefront models and evanescent waves, which decay exponentially beyond the \ac{RIS} surface but contribute significantly to subwavelength focusing. Thus, configurations should use spatially dependent formulations rather than only angular steering. Accurate modeling across both domains requires careful modeling of wave propagation and scattering. At $3.5~\text{GHz}$, corresponding to a wavelength of approximately $0.0857~\text{m}$, the behavior of the RIS changes significantly as the measurement distance shifts from the near-field at 2~m to the far-field at 10~m. Therefore, we propose and illustrate the \ac{MF-RIS} phase synthesis algorithm in Fig. \ref{fig:alg}, showing that the codebook generation process differs fundamentally between the far-field and near-field domains.

\subsection{Near-Field Codebook}

When the transmitter or receiver lies within the Fraunhofer distance
\(<2D^2/\lambda\) (with \(D\) the largest \ac{RIS} dimension), the impinging and reflected wavefronts must be modeled as spherical rather than plane waves to maximize the \ac{RIS} beamforming gain. This relationship is conventionally modeled through the Green’s function–based \ac{3D} coupling among the \ac{BS}, the \ac{RIS} elements \((m,n)\), and the \ac{UE}, as shown below:
\begin{equation}
\mathbf{E}(\mathbf r) \;=\;
\sum_{m=1}^{M}\sum_{n=1}^{N}
\Gamma_{m,n}\,
\frac{e^{-j k \big(\|\mathbf r_{m,n}^{\text{BS--RIS}}\|+\|\mathbf r_{m,n}^{\text{RIS--UE}}\|\big)}}
{\|\mathbf r_{m,n}^{\text{BS--RIS}}\|\,\|\mathbf r_{m,n}^{\text{RIS--UE}}\|},
\quad k=\tfrac{2\pi}{\lambda},
\label{eq:GF}
\end{equation}
where \(\Gamma_{m,n}\) is the (generally complex) reflection coefficient of the
\((m,n)\)-th RIS element, \(\mathbf r_{m,n}^{\text{BS--RIS}}\) and
\(\mathbf r_{m,n}^{\text{RIS--UE}}\) are the BS–to–element and element–to–\ac{UE} distance vectors, respectively, \(\lambda\) is the wavelength, and
\(\|\cdot\|\) denotes the Euclidean norm.

By appropriately adjusting \(\Gamma_{m,n}\), the superposition of the scattered
fields can be engineered to focus within the near-field region. Let
\(\Psi_{m,n}\) denote the programmed phase of the \((m,n)\)-th unit cell.
To ensure that all contributions in \eqref{eq:GF} add in phase at the focal
point, the \ac{RIS} should compensate both the propagation phase and the residual quadratic (Fresnel) curvature from the two hops. In other words, the phase delays due to propagation from \ac{BS} to $(n,m)$ and from $(n,m)$ to \ac{UE} should be exactly compensated by the \ac{MF-RIS} element’s added phase and the \ac{MF-RIS} focal point:
\begin{align}
&\Psi_{m,n}^{\text{NF}}
= \operatorname{mod}\!\Bigg\{
-k\!\left(\big\|\mathbf r_{m,n}^{\text{BS--RIS}}\big\|
           +\big\|\mathbf r_{m,n}^{\text{RIS--UE}}\big\|\right) \nonumber\\
&\;\;-\;\frac{k}{2}\,\big(x_{m,n}^2+y_{m,n}^2\big)\!
\left(
\frac{\cos^{2}\theta_{i}}{\big\|\mathbf r_{m,n}^{\text{BS--RIS}}\big\|}
+\frac{\cos^{2}\theta_{r}}{\big\|\mathbf r_{m,n}^{\text{RIS--UE}}\big\|}
\right)
\;,\; 2\pi \Bigg\},
\label{eq:NF}
\end{align}
where $x_{m,n}$ and $y_{m,n}$ are the \ac{2D} coordinates of the $(m,n)$'th RIS element, and \(\theta_{i}\) and \(\theta_{r}\) are the incidence and reflection angles (with respect to the \ac{RIS} normal). Note that the conversion from spherical to Cartesian coordinates is required to realize the reflection in the desired reflection angle $\left(\theta_r, \phi_r\right)$.

\begin{figure*}[!t]
\centering

    \begin{turn}{0}
    \includegraphics[width=2\columnwidth]{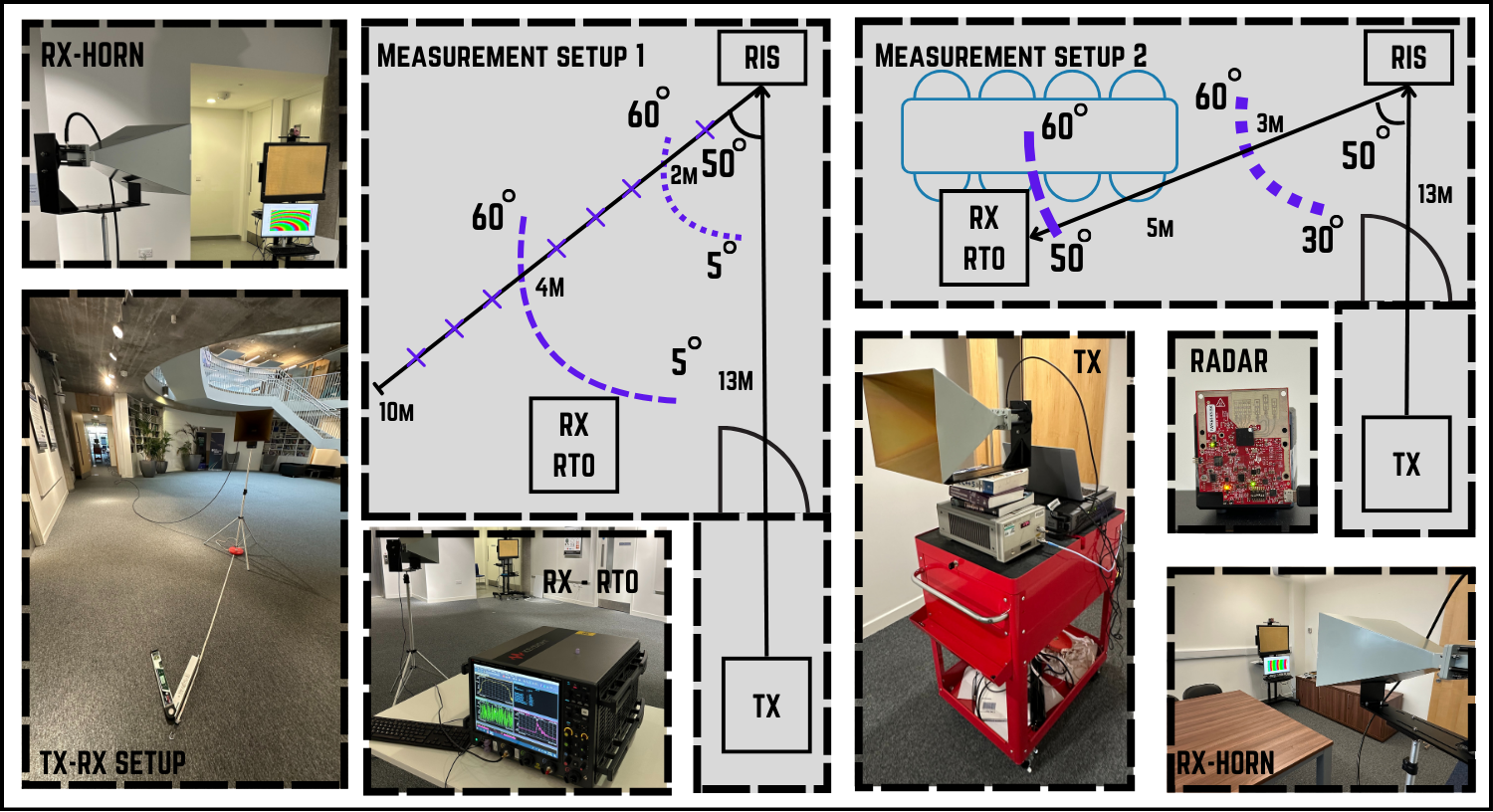} 
    \end{turn}

\caption{Static measurement setup: \ac{RIS} reflecting the signal from a \ac{TX} located in a far-field towards a \ac{RX} located at varied distances in $5^\circ$ angle and 1 m steps. A large indoor space and a small meeting room were evaluated.}
\label{setup}
\end{figure*}

\begin{figure}[!t]
\centering

    \begin{minipage}{1\columnwidth} 
        \centering
        \begin{subfigure}[b]{0.48\textwidth}
            \centering
            \includegraphics[width=\textwidth]{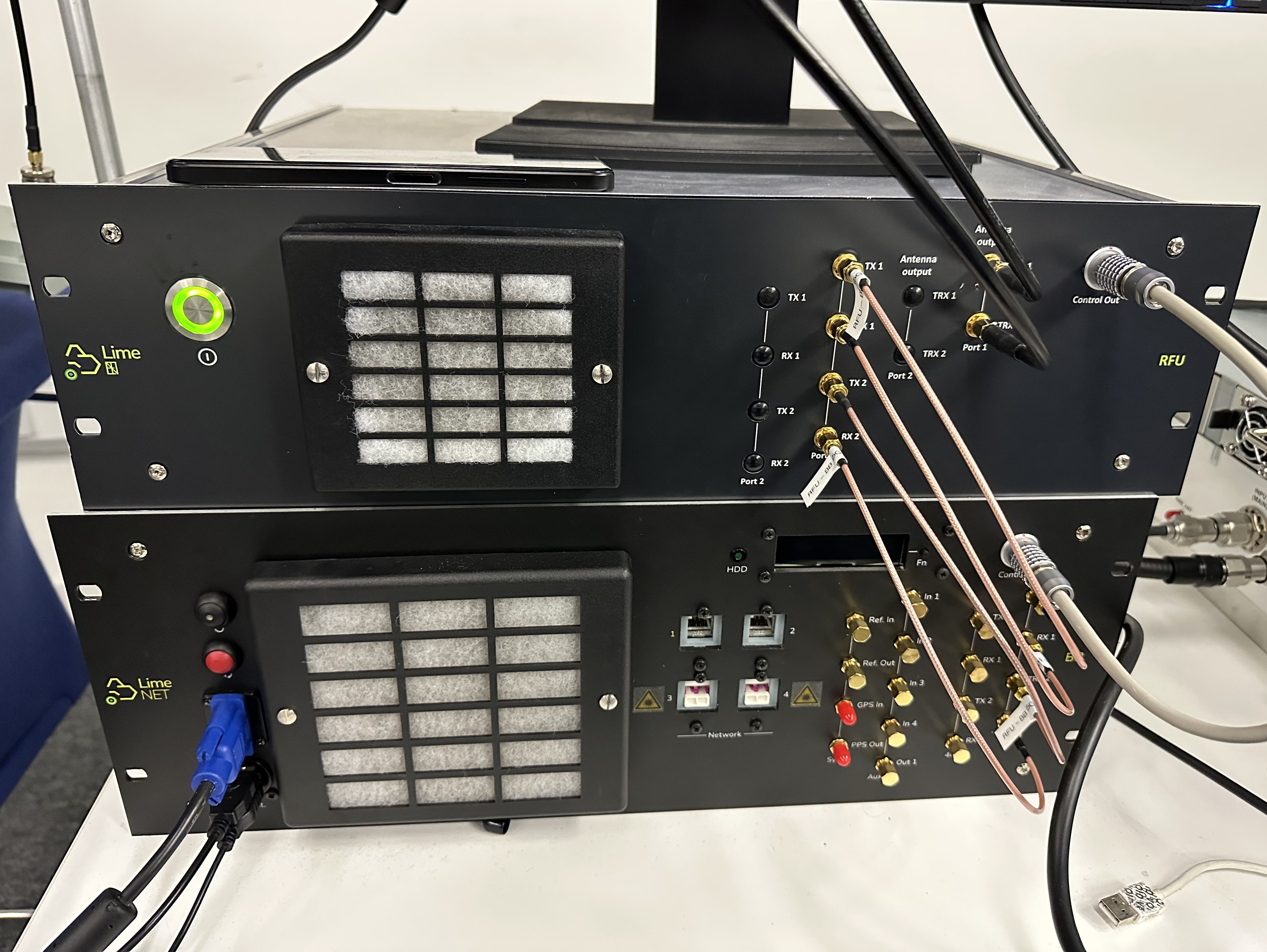} 
            \caption{LimeNET BS}
        \end{subfigure}
        \hfill
        \begin{subfigure}[b]{0.48\textwidth}
            \centering
            \includegraphics[width=\textwidth]{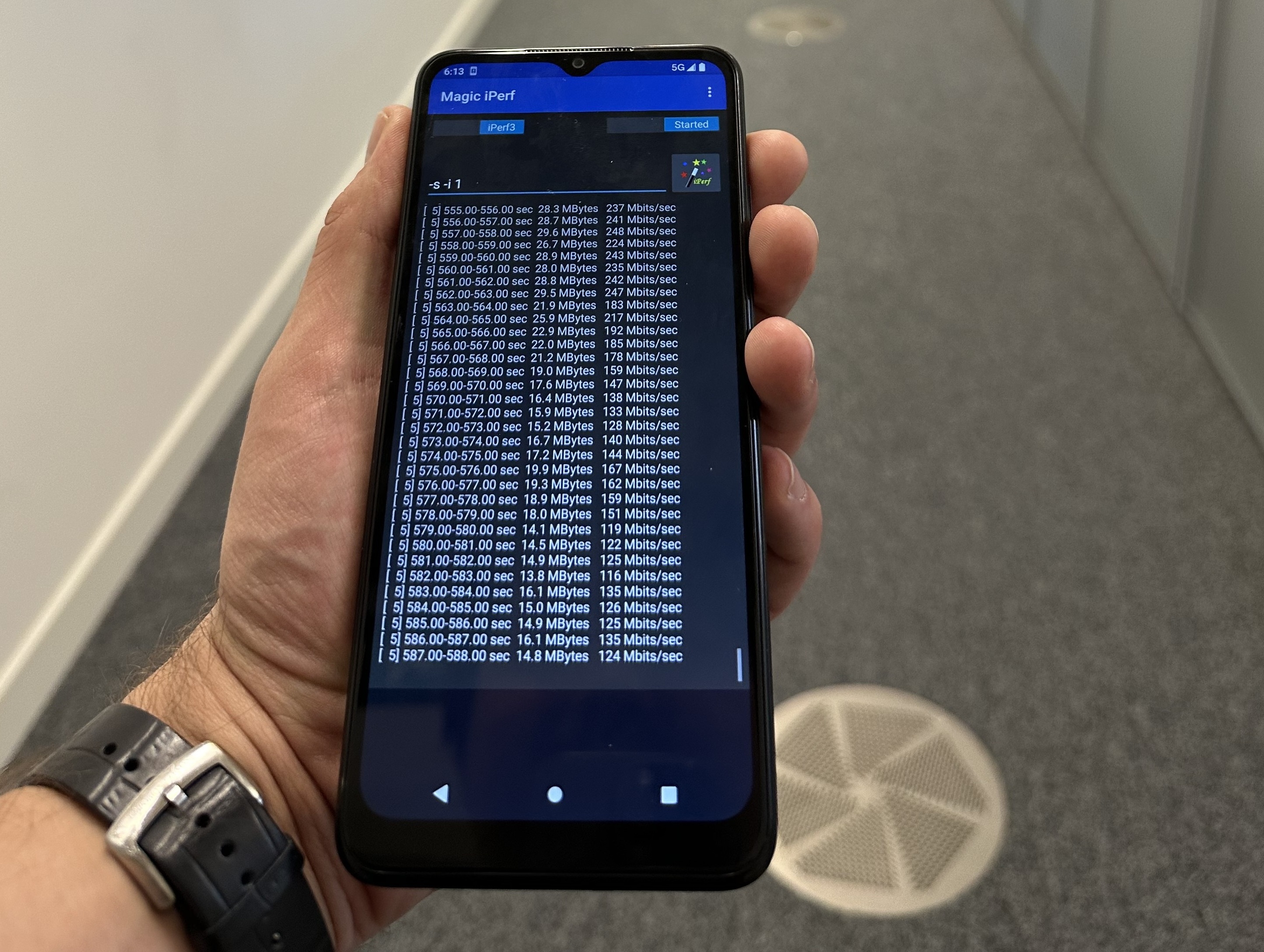} 
            \caption{UE measurements}
        \end{subfigure}
    \end{minipage}

\caption{Dynamic \ac{UE} measurement setup: a) the LimeNET 2x2 \ac{MIMO} system b) iPerf test function performed on a 4x4 \ac{MIMO} smartphone to obtain the throughput measurement}
\label{Lime}
\end{figure}

\subsection{Far-Field Codebook}
As the \ac{UE} moves further away from \ac{MF-RIS}, such as to 10 meters or beyond, the spherical curvature of the wavefronts diminishes and the far-field assumption becomes valid, making the contributions from distance-based terms negligible. In this case, a linear phase gradient suffices for \ac{RIS} beamsteering:
\begin{align}
\Psi_{m,n}^{\text{FF}}
= & \operatorname{mod}\Bigg\{
   -k \Big(
   x_{m,n}\,\sin\theta_r \cos\phi_r  \\
& + y_{m,n}\,\sin\theta_r \sin\phi_r
   \Big),\, 2\pi
   \Bigg\}, \notag
\end{align}
This model assumes a planar incident wavefront with $\left(\theta_i,\phi_i\right) = (0^\circ,0^\circ)$ and a uniform array structure.

To illustrate these principles, Fig.~\ref{near} shows the \ac{RIS} configuration for $\theta=40^\circ$, $\phi=90^\circ$ at 2 meters, the \ac{RIS} near-field beam pattern, and the corresponding peak of the \ac{PDP} for \ac{RIS} on and off cases. Fig.~\ref{far} presents the same comparison using a linear phase gradient codebook in the far-field regime at 10 meters. \ac{PDP} peak shows 17.5 dB $/$ 7.5 dB improvement with \ac{RIS} in the near-field and the far-field, respectively.

\begin{table}[!t]
\renewcommand{\arraystretch}{1.3}
\captionsetup{justification=centering}\captionsetup{ labelfont={sc,footnotesize}}
\caption{\footnotesize\sc {RIS Specification}}
\label{table_RIS_spec}
\centering
\begin{tabular}{|l|l|}
\hline
\textbf{Operating Frequency} & 3.5 GHz (centre) \\
\hline
\textbf{Bandwidth} & 800 MHz \\
\hline
\textbf{Type} & Reflective \\
\hline
\textbf{Polarization} & Dual-polarized \\
\hline
\textbf{Size} & 570 mm $\times$ 420 mm \\
\hline
\textbf{Number of unit cells} & 50 $\times$ 37 with $\lambda/8$ spacing (1,850 in total)\\
\hline
\textbf{Unit cell control} & 4-bit digital \\
\hline
\textbf{Switching element} & Varactor diode \\
\hline
\textbf{Power consumption} & 2.2~W (RIS) + 2~W (Radar) + 10~W (SBC)  \\
\hline
\end{tabular}
\end{table}

\setlength{\fboxrule}{1.5pt} 

\setlength{\fboxsep}{5pt} 

\section{Channel Measurement Setup}\label{ch6}
All measurements were performed in the \ac{ICS}, \ac{5G} / \ac{6G} Innovation Centre, University of Surrey. Measurement setups used in this paper are shown in Fig. \ref{setup} and consider two of the typical indoor \ac{RIS} use-case scenarios: a) a large auditorium hall $\left(14.6\times10.8~\mathrm{m}^2\right)$, and b) a meeting room $\left(7.3\times4.3~\mathrm{m}^2\right)$. In either case, the \ac{LoS} path between the \ac{BS} and the \ac{UE} is blocked, however, the \ac{NLoS} path exists, which was confirmed with measurement. 
\subsection{Device Under Test}
All measurements were carried out using the SMV2201-040LF varactor diode-based 4-bit \ac{RIS}, which has a centre frequency of 3.5 GHz and the other characteristics listed in Table \ref{table_RIS_spec}. It utilizes matrix-based control of 1,850 unit cell elements, arranged in 37 rows and 50 columns, resulting in a total aperture size of 420 x 570 mm. 
This \ac{RIS} forms a part of the \ac{MF-RIS} system with sensing capabilities, where the 60 GHz \ac{mmWave} radar IWR6843ISK by \Ac{TI} is used for implementing the custom UE tracking algorithm on the Seedstudio ODYSSEY X86J4125800 \ac{SBC} to dynamically update the \ac{MF-RIS} codebook.


\subsection{Static Measurement}
In this setup, the \ac{RX} remained static, and the measurements were taken from designated positions rather than in real-time. The transmitted signals were generated using a Keysight M8195A waveform generator at the \ac{TX}, amplified by a 21 dBm gain \ac{LNA}. At the \ac{RX}, a Keysight UXR0334B Infiniium \ac{RTO} captured the resulting complex channel responses. Both the \ac{TX} and \ac{RX} employed identical A-INFOMW LB-229-20-C-SF horn antennas, each providing 20 dBi gain. A total of 1024 symbols were transmitted at a rate of 800 Msps, corresponding to a \ac{FFT} length of 1024 and an operational bandwidth of 800 MHz, giving a delay bin resolution of 1.25 ns and a corresponding distance resolution of 0.375 m. The \ac{RTO} was calibrated at the cable ends to remove connector losses and non-linearities of the \ac{LNA}. Each measurement was repeated three times to ensure repeatability.
The measurements were used to obtain the time-domain channel response, with each response averaged over 50 individual realizations. From these responses, the \ac{PDP} and \ac{RMS} delay spread were derived at each measurement point to characterize the static propagation environment and \acp{MPC}. Both the transmitting antenna and the receiving antenna were positioned at a height of 1.45 m above the floor, aligning them with the centre of the \ac{RIS}. The position of \ac{TX} and \ac{RIS} was fixed, whereas the position of the \ac{RX} was adjusted between near-field (2~m) and far-field (10~m) distances for different measurements. Measurements were taken at angles between 5$^\circ$ to 60$^\circ$ every 5$^\circ$ steps for further verification of the \acp{RIS} performance. The orientation of the \ac{RX} relative to the \ac{RIS} was configured to suit the reflection profile of the \ac{RIS}. During the measurements, there was no human presence or moving objects, ensuring that the channel remains time-invariant. Data from each measurement session was recorded on the \ac{RTO}, subsequently transferred, and analyzed using MATLAB.

\subsection{Dynamic Measurement}
To perform the channel measurement of a dynamically moving \ac{UE}, the LimeNET \ac{5G} \ac{NR} 2x2 \ac{MIMO} system shown in Fig. \ref{Lime} was used in a single user configuration. The system produces a 3.5 GHz modulated waveform with an initial power of 10 dBm and 30 dBm connected to two A-INFOMW LB-229-20-C-SF standard 20~dBi gain horn antennas, both aiming the wavefront towards the \ac{MF-RIS}. On the receiving side, the 4x4 \ac{MIMO} smartphone was wirelessly attached to the LimeNET network and equipped with iPerf software to monitor the channel quality characteristics. The smartphone position was moved dynamically within the \ac{MF-RIS} near-field at 2~m and 4~m distances across 5-60$^\circ$ angular range with a constant velocity, while observing throughput as a system evaluation metric.




  



\section{Results and Analysis}\label{ch7}
\par
\begin{figure}[!t]
\centering

\begin{subfigure}{.96\linewidth}
  \includegraphics[width=\linewidth]{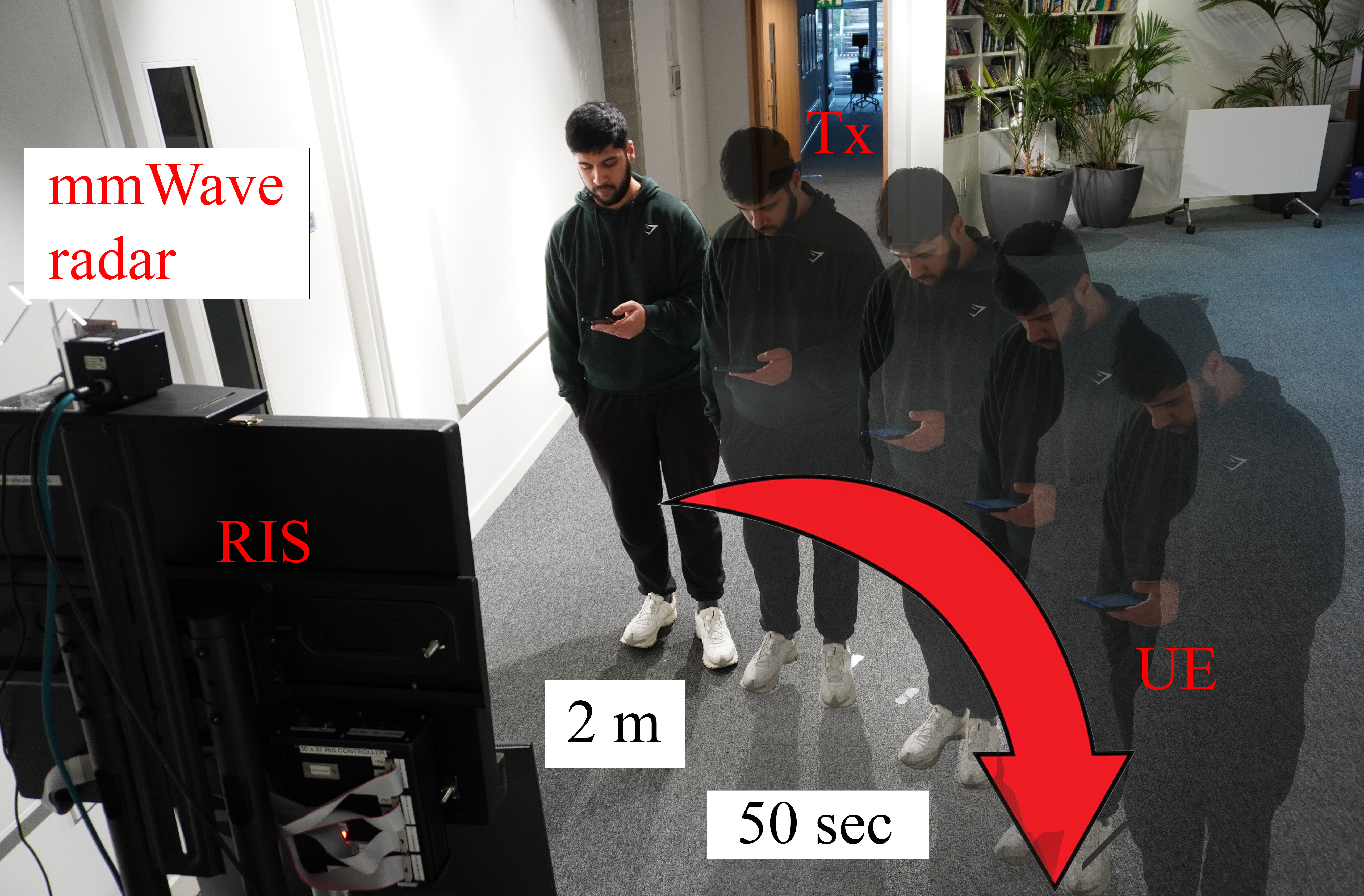}
  \caption{}
\end{subfigure}\hfill

\begin{subfigure}{.495\linewidth}
  \includegraphics[width=\linewidth]{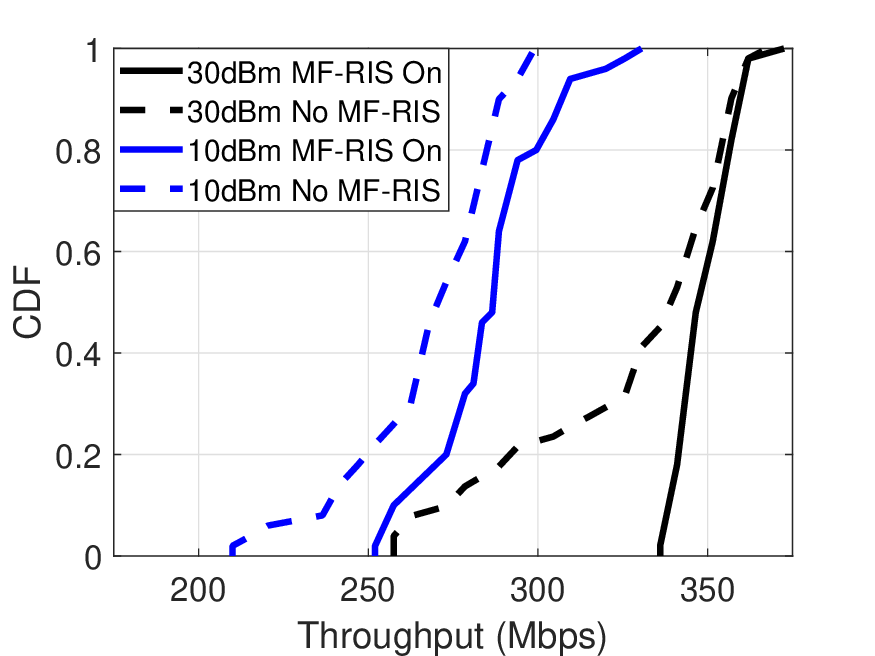}
  \caption{}
\end{subfigure}\hfill
\begin{subfigure}{.495\linewidth}
  \includegraphics[width=\linewidth]{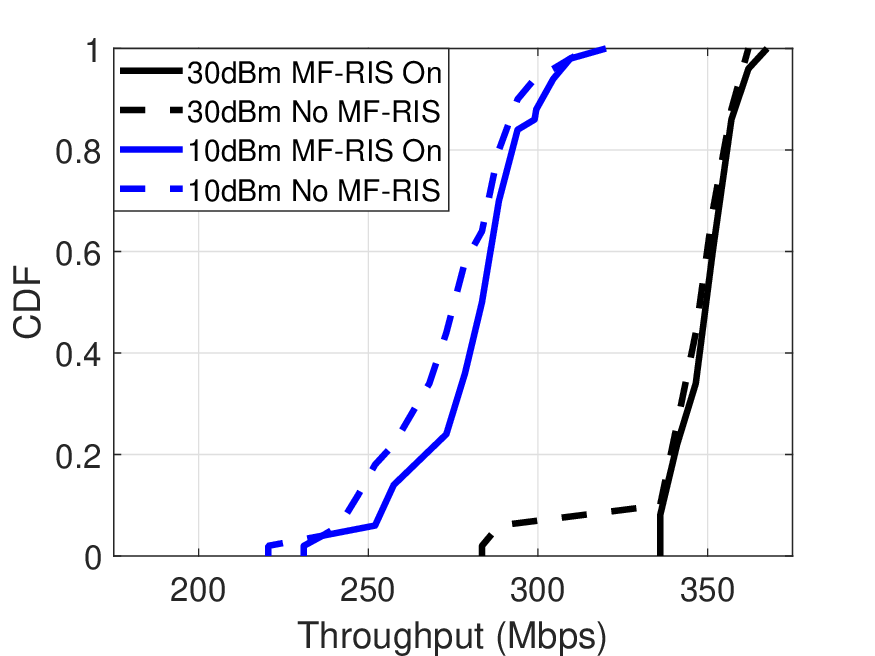}
  \caption{}
\end{subfigure}

\caption{System throughput measurements in the near-field: a) \ac{MF-RIS}-enabled \ac{RCC} system setup at 2~m, b) CDF of the channel throughput measurement at 2~m, and c) 4~m distances with \ac{MF-RIS} (solid) and without \ac{MF-RIS} (dashed).}
\label{Throughput}
\end{figure}

This section presents the results obtained from the empirical data collected from the measurement. Large-scale and small-scale channel parameters are summarized in Tables III and IV, which were obtained from the two static measurement setups described previously\footnotemark[3]. Note that clustering distributions are based on normalized \ac{PDP}, and absolute values should be obtained by applying the path loss that was obtained at the center frequency corresponding to the \ac{PDP} peak. At the end of the section, the dynamic \ac{UE} throughput measurement results on the LimeNET \ac{MIMO} system are described and illustrated in Fig. \ref{Throughput}. Note that the calculated theoretical maximum throughput for this 2$\times$2 \ac{MIMO} system is 571.1 Mbps.


\begin{table*}[!t]
\centering
\captionsetup{justification=centering, labelfont={sc,footnotesize}}
\caption{\footnotesize\scshape Measurement Setup 1 — Variable RIS-UE Reflection Angle (5$^\circ$–60$^\circ$) and RIS-UE Distance (1–10\,m).}
\label{tab:ms1_combined}
\footnotesize
\setlength{\tabcolsep}{4pt}
\renewcommand{\arraystretch}{1.15}

\subfloat[Measured BS-UE Channel Parameters without RIS at a Fixed  2~m Distance (Angular Sweep)]{
\resizebox{\textwidth}{!}{%
\begin{tabular}{lcccccccccccc}
\toprule
Deg. & $5^\circ$ & $10^\circ$ & $15^\circ$ & $20^\circ$ & $25^\circ$ & $30^\circ$ & $35^\circ$ & $40^\circ$ & $45^\circ$ & $50^\circ$ & $55^\circ$ & $60^\circ$ \\
\midrule
$\text{PL}^{\text{BS-UE}}_{\text{CI}} (\text{dB})$
& 107.5 & 103.3 & 112.9 & 105.8 & 94.5 & 110.8 & 97.8 & 102.8 & 108.2 & 98.6 & 107.7 & 107.7 \\
$\sigma_\text{SF} (\text{db})$    &4.6 &5.4 &6.3 &4.6 &7.7 &4.9 &4.9 &4.4 &5.7 &4.4 &4.9 &4.9 \\
$K_\text{R}(\text{dB})$                           & -0.4  & 3.5  & -1.8 &  -0.6 & -3.6  & -1.4  & -1.4  & -2.1  & 0.5  &  1.6 &  2 & 1.9  \\
$\tau_{\text{rms}}(\text{ns})$                           &  20.4 &  18.3 &  20.2 &  23.2 & 27.7 & 21 & 23.5 & 25.9  & 23.1  & 24.5 & 26.5 & 26.1 \\
\midrule
$W_{\text{o}}$                &  5    &  6    &   7   &   7   &   8   &  5    &  4    &   6   &  7   &   8   &  8    &  5    \\
$DS_{o=1} (\text{ns})$                          & 1.6     &   1.2   &  1.84    &  2.35    &   1.52   &  0.83    &  0.96    &  1.85    & 1.5     &   1.34   &  1.1    &  0.78    \\
$\zeta^2_{o=1} (\text{dB})$                   &  2.41    &  6.9    & 2.8    &   1.6   &  4    & 2.3     &3.9      & 5.4     &   2.6   &  2.9    & 4.3     &  2.4    \\
$r^\tau_{o=1}$                      &  \shortstack{$0.09\,e^{+0.048\tau}$ \\ + $10^{-3}$}    &   \shortstack{$6.9\,e^{-1.23\tau}$ \\ + $10^{-3}$}   &  \shortstack{$3.3\,e^{-0.64\tau}$ \\ + $10^{-3}$}    & \shortstack{$0.05\,e^{-0.13\tau}$ \\ + $10^{-3}$}      &  \shortstack{$6.8\,e^{-1.1\tau}$ \\ + $10^{-2}$}     & \shortstack{$87.8\,e^{-1.8\tau}$ \\ + $3.2*10^{-3}$}       &    \shortstack{$0.07\,e^{-0.27\tau}$ \\ + $10^{-3}$}   &  \shortstack{$0.59\,e^{-0.56\tau}$ \\ + $10^{-3}$}      &  \shortstack{$4.71\,e^{-2.31\tau}$ \\ + $10^{-3}$}     &   \shortstack{$2.55\,e^{-1.21\tau}$ \\ + $10^{-3}$}    &   \shortstack{$7.1\,e^{-1.1\tau}$ \\ + $10^{-3}$}    &   \shortstack{$7.1\,e^{-1.1\tau}$ \\ + $10^{-3}$}    \\
\bottomrule
\end{tabular}}}
\\[1ex]
\subfloat[Measured BS-RIS-UE Channel Parameters with RIS at a Fixed 2~m Distance (Angular Sweep)]{
\resizebox{\textwidth}{!}{%
\begin{tabular}{lcccccccccccc}
\toprule
Deg. & $5^\circ$ & $10^\circ$ & $15^\circ$ & $20^\circ$ & $25^\circ$ & $30^\circ$ & $35^\circ$ & $40^\circ$ & $45^\circ$ & $50^\circ$ & $55^\circ$ & $60^\circ$ \\
\midrule
$\text{PL}^{\text{BS-RIS-UE}} _{\text{CI}}(\text{dB})$
& 82.5 & 79.6 & 79.5 & 84.5 & 81.2 & 79.6 & 82.7 & 83.1 & 79.8 & 83.1 & 82.5 & 83.1 \\
$\sigma_\text{SF} (\text{dB})$     &5.4 &5.5 &4.5 &3.8 &4.7 &3.7 &4.4 &4.6 &4.5 &6.9 &7.7 &4.9 \\
$K_\text{R}(\text{dB})$                           & 6  &  6.4 & 5.7  &  11.3 & 6.8  & 7.5  & 12.1  &  12.6 & 8  &  10.6 & 9 & 10.3\\
$\tau_{\text{rms}} (\text{ns})$                           & 17  & 15.7  & 18.8  & 15.2  & 18.9  &  19.1 &  14 & 11.5  & 16.6  & 13.1 &  15.4 & 14.8 \\
\midrule
$I_{\text{q}}$            & 15     &   15   &  18    &  18    &  10    & 12     &    7  &     17 &  9    &  6    &  7    &  13    \\
$DS_{q=1} (\text{ns})$                      &2.7      &  2.4    &  5    &   0.5   &  1.8    &  3.2    &    2.8  &  2.8    &  1.4    & 1.1     &   0.73   &    1.5  \\
$\zeta^2_{q=1} (\text{dB})$               &   3.8   &  3.4    &  3    &   2.75   & 2.5     &   1.75   &  4.8    & 2.1     & 2.2     & 2.7     &   1   &    2  \\
$r^\tau_{q=1}$                  &
\shortstack{$1.8\,e^{-1.14\tau}$ \\ + $10^{-3}$} &
\shortstack{$10.3\,e^{-1.76\tau}$ \\ + $10^{-3}$} &
\shortstack{$2.36\,e^{-1.2\tau}$ \\ + $10^{-3}$} &
\shortstack{$2.26\,e^{-1.46\tau}$ \\ + $3.2*10^{-4}$} &
\shortstack{$0.07\,e^{-0.48\tau}$ \\ + $3.2*10^{-4}$} &
\shortstack{$0.14\,e^{-0.75\tau}$ \\ + $10^{-3}$} &
\shortstack{$3.94\,e^{-1.6\tau}$ \\ + $10^{-3}$} &
\shortstack{$10\,e^{-1.9\tau}$ \\ + $10^{-3}$} &
\shortstack{$0.73\,e^{-0.89\tau}$ \\ + $10^{-3}$} &
\shortstack{$12.76\,e^{-1.9\tau}$ \\ + $2*10^{-3}$} &
\shortstack{$7.56\,e^{-1.87\tau}$ \\ + $10^{-3}$} &
\shortstack{$3.15\,e^{-1.55\tau}$ \\ + $3.2*10^{-4}$} \\
\bottomrule
\end{tabular}}}

\vspace{1ex}
\renewcommand*{\arraystretch}{1.25}
\subfloat[Measured BS-UE Channel Parameters without RIS at a Fixed 50$^\circ$ Angle (Distance Sweep)]{
\resizebox{\textwidth}{!}{%
\begin{tabular}{l*{10}{c}}
\toprule
Dist. & 1m  & 2m  & 3m  & 4m  & 5m  & 6m  & 7m  & 8m  & 9m  & 10m \\
\midrule
$\text{PL}^{\text{BS-UE}}_{\text{CI}}(\text{dB})$
& 92.2 & 98.6 & 108.9 & 110.3 & 111.3 & 112.8 & 118.6 & 115.9 & 118.4 & 118.9 \\
$\sigma_\text{SF}(\text{dB})$     &5 &4.3 &5.2 &6.7 &6.7 &5.7 &5.7 &5.8 &5.7 &6.3 \\
$K_\text{R} (\text{dB})$                           & -3.8  &  0.6 & -5  &  -4.6 & -0.6  & -0.3  &  -1.6 & -2.8  & -3.7 & -8.1  \\
$\tau_{\text{rms}} (\text{ns})$                           & 26.1  & 24.5  & 27.8  & 30 & 29  &  26.3 & 23.7 & 27.2 & 27.3  &  26.5  \\
\midrule
$W_{\text{o}}$                & 9 &  6&  6& 4& 3 & 3 & 7 & 11 & 3 & 5  \\
$DS_{o=1} (\text{ns})$                          &  2.4& 1.12 & 1.37 & 1.15 & 0.87 & 1 & 3 & 3 & 1 &  1.34 \\
$\zeta^2_{o=1} (\text{dB})$                     &  3.8&3.2  & 10.9 & 5.7 & 4.4 & 1.1 & 3.9 & 3.6 & 0.25 & 5.2 \\
$r^\tau_{o=1}$                      &\shortstack{$0.51\,e^{-0.29\tau}$ \\ + $10^{-3}$}    & \shortstack{$1.2\,e^{-1.21\tau}$ \\ + $10^{-3}$}  & \shortstack{$0.93\,e^{-0.21\tau}$ \\ + $10^{-3}$}  & \shortstack{$0.03\,e^{+0.04\tau}$ \\ + $10^{-3}$}  & \shortstack{$2.88\,e^{-1.35\tau}$ \\ + $10^{-3}$}  &\shortstack{$0.44\,e^{-0.24\tau}$ \\ + $10^{-3}$}    & \shortstack{$0.85\,e^{-0.16\tau}$ \\ + $10^{-3}$}  & \shortstack{$1.1\,e^{-0.32\tau}$ \\ + $10^{-3}$} & \shortstack{$0.49\,e^{-0.06\tau}$ \\ + $10^{-3}$}  & \shortstack{$0.38\,e^{+0.14\tau}$ \\ + $10^{-3}$}  \\
\bottomrule
\end{tabular}}}
\\[1ex]
\subfloat[Measured BS-RIS-UE Channel Parameters with RIS at a Fixed 50$^\circ$ Angle (Distance Sweep) ]{
\resizebox{\textwidth}{!}{%
\begin{tabular}{l*{10}{c}}
\toprule
Dist. & 1m  & 2m & 3m & 4m & 5m & 6m  & 7m  & 8m  & 9m  & 10m \\
\midrule
$\text{PL}^{\text{BS-UE}}_{\text{CI}} (\text{dB})$
& 76.8 & 83.1 & 89.5 & 88.4 & 94.8 & 95.1 & 95.6 & 99 & 93 & 94.9 \\
$\sigma_\text{SF} (\text{dB})$  & 5.8 &6.9 &7 &6 &3.8 &3.4 &4.7 &5.4 &5 &4.1 \\
$K_\text{R} (\text{dB})$                           & 9.2  &  10.6 &  4.1 & 5.6  & 6.7  & 4.1  &  3.5 &  1.7 & 2.2  &  0.5  \\
$\tau_{\text{rms}} (\text{ns})$                           & 14.8  & 13.1  & 21.8  &  22.2 & 23.3  &  21 & 20.5  &  24.7 & 25  &  27  \\
\midrule
$I_{\text{q}}$                & 6   & 6  & 11   & 6   & 6   & 8   & 9   & 8   & 14   & 5  \\
$DS_{q=1} (\text{ns})$                          &  0.76&   0.62 &   2   & 1   & 2.4   &2.5    & 3.14   & 2.4   & 4.1   &1.73  \\
$\zeta^2_{q=1} (\text{dB})$                   & 1.63   &2.9    & 1.9   & 3.7   & 6.41   & 3.3   & 3.85   & 3.79   &2.8    &  0.62\\
$r^\tau_{q=1}$                      &\shortstack{$0.43\,e^{-1.17\tau}$ \\ + $10^{-4}$}    &\shortstack{$4.6\,e^{-2\tau}$ \\ + $10^{-3}$}    &  \shortstack{$1.25\,e^{-1.4\tau}$ \\ + $10^{-3}$}&   \shortstack{$2\,e^{-1.8\tau}$ \\ + $10^{-3}$} &   \shortstack{$0.03\,e^{-0.01\tau}$ \\ + $10^{-3}$}   & \shortstack{$0.07\,e^{-0.06\tau}$ \\ + $10^{-3}$}   & \shortstack{$0.23\,e^{-0.24\tau}$ \\ + $10^{-3}$}   & \shortstack{$0.14\,e^{-0.11\tau}$ \\ + $10^{-3}$}   & \shortstack{$0.22\,e^{-0.21\tau}$ \\ + $10^{-3}$}   & \shortstack{$0.18\,e^{-0.1\tau}$ \\ + $10^{-3}$}      \\
\bottomrule
\end{tabular}}}
\end{table*}

\begin{figure*}[!t]
\centering

\begin{subfigure}[b]{0.24\linewidth}
    \centering
    \includegraphics[width=\linewidth]{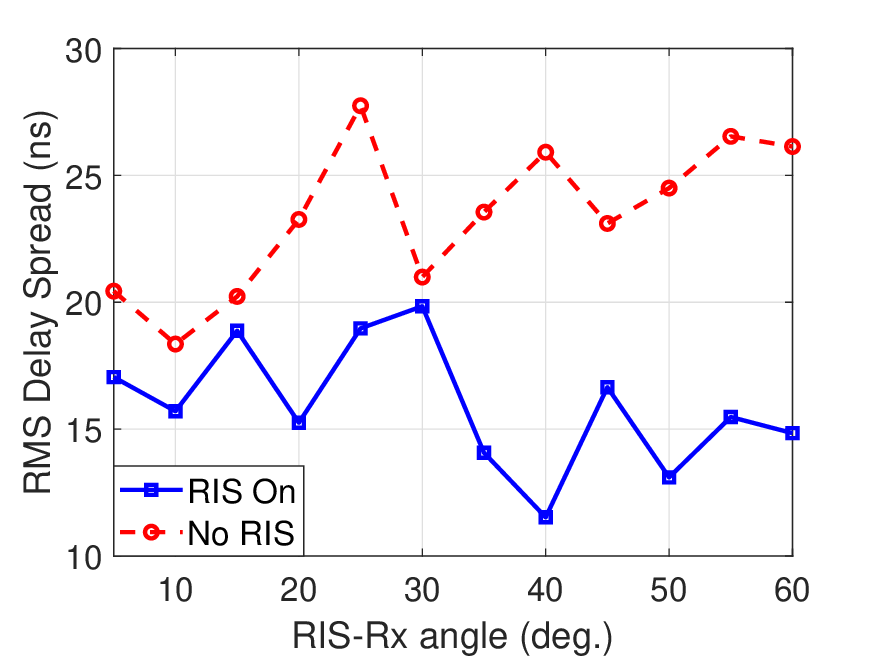}
    \caption{}
\end{subfigure}
\hfill
\begin{subfigure}[b]{0.24\linewidth}
    \centering
    \includegraphics[width=\linewidth]{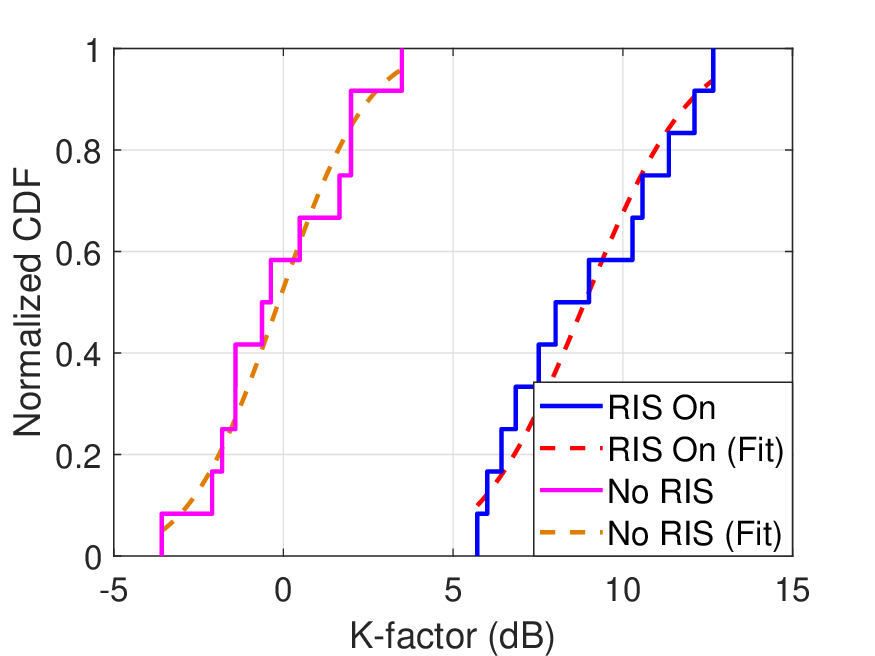}
    \caption{}
\end{subfigure}
\hfill
\begin{subfigure}[b]{0.24\linewidth}
    \centering
    \includegraphics[width=\linewidth]{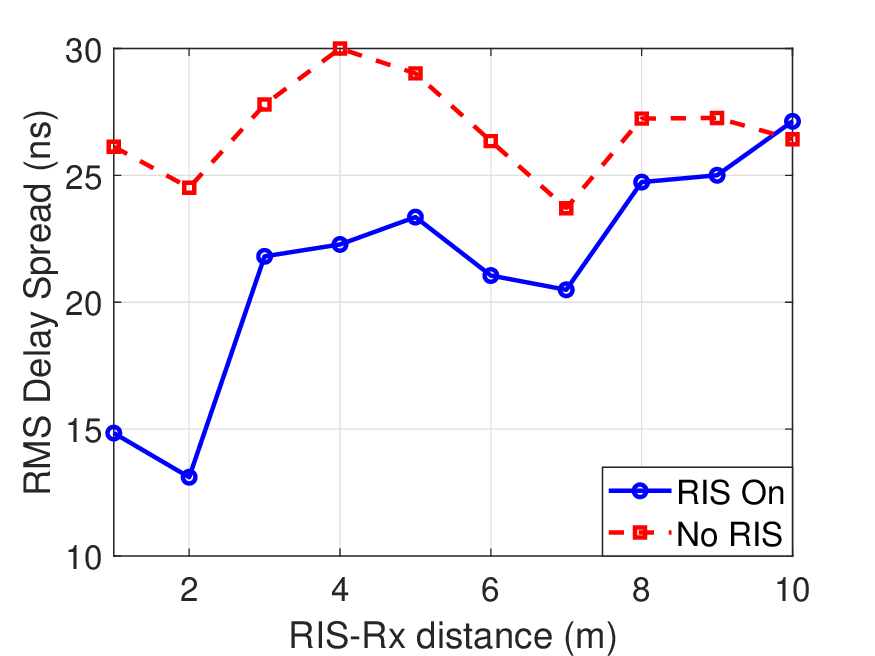}
    \caption{}
\end{subfigure}
\hfill
\begin{subfigure}[b]{0.24\linewidth}
    \centering
    \includegraphics[width=\linewidth]{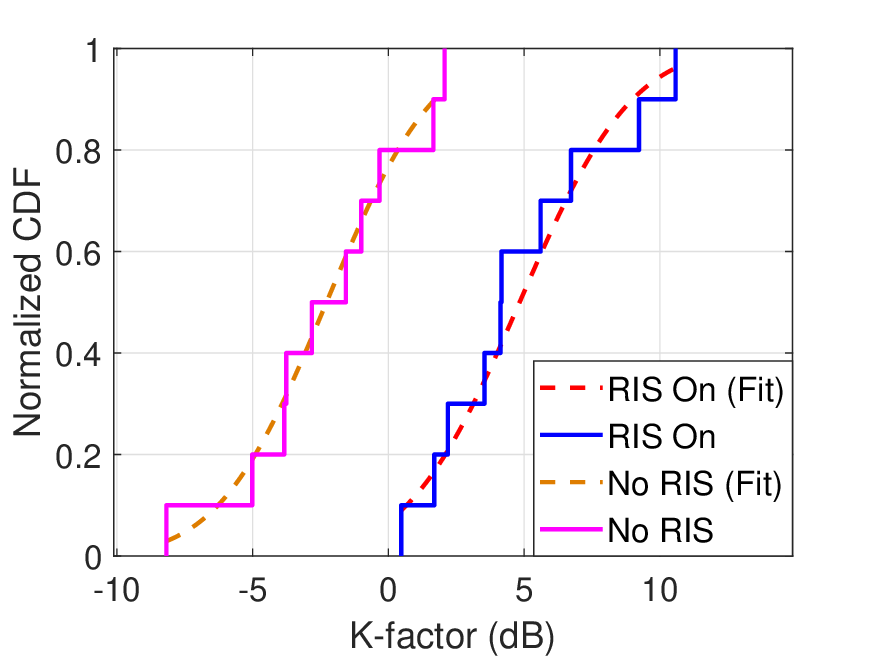}
    \caption{}
\end{subfigure}

\caption{\ac{RMS} delay spread and K-factor analysis of raw \acp{PDP}: 1. Angular measurements between 5$^\circ$-60$^\circ$ at a fixed RIS-Rx distance of 2m: a) RMS delay spread, b) CDF of the K-factor and the normal fit. 2. Spatial measurements for RIS-Rx distances between 1-10m with fixed 50$^\circ$ reflection angle: c) RMS delay spread, d) CDF of the K-factor and the normal fit.}
\label{fig:ds-kf}
\end{figure*}

\begin{table*}[!htp]
\centering
\captionsetup{justification=centering, labelfont={sc,footnotesize}} 
\caption{\footnotesize\scshape Measurement Setup 2: Variable RIS-UE Angle (30$^\circ$-60$^\circ$) and RIS-UE Distance (3~m and 5~m).}

\renewcommand*{\arraystretch}{1.25}


\begin{subtable}{\textwidth}
\centering
\subfloat[Measured BS-UE Channel Parameters without RIS]{
\resizebox{\textwidth}{!}{%
\begin{tabular}{l*{10}{c}}
\toprule
Deg.~/~Dist. & 30$^\circ$~/~3m  & 35$^\circ$~/~3m  & 40$^\circ$~/~3m  & 45$^\circ$~/~3m  & 50$^\circ$~/~3m & 55$^\circ$~/~3m  & 60$^\circ$~/~3m  & 50$^\circ$~/~5m  & 55$^\circ$~/~5m & 60$^\circ$~/~5m \\
\midrule
$\text{PL}^{\text{BS-UE}}_{\text{CI}}(\text{dB})$
& 96.0 & 99.5 & 88.4 & 92.8 & 92.6 & 98.0 & 94.6 & 95.8 & 104.4 & 95.3 \\
$\sigma_\text{SF} (\text{dB})$                          & 6.3 &  7 & 7.1  & 7.7  &  7.3 &  6.1 & 6.2  & 6.8  & 6.9  & 5.9    \\
$K_\text{R} (\text{dB})$                          & 5.3  & 4.2  &  5.2 & 4  &  5.5 &  0.1 & 8.4  & 2.2  & 6.1  &  5.8  \\
$\tau_{\text{rms}} (\text{ns})$                          & 22 &  29.6 & 26  & 28  & 31.4  &  33.4 & 34.8  &   24.9 &  19.4 &  30.1  \\
\midrule
$W_{\text{o}}$                 & 9 & 11  &  8 &  11 & 11  & 11  &  12 &  10 &  12 &  11  \\
$DS_\text{o=1} (\text{ns})$                           & 14 &  14.1 & 10.9  & 9  & 12.3  & 13.1  & 17.5  & 11.1  & 14.4  & 14.2    \\
$\zeta^2_\text{o=1} (\text{dB})$                   & 6  & 4  & 4.4  &  3.5 &  4.8 & 5.1  & 3.4  &  3.1 & 6.3  &  4   \\
$r^\tau_\text{o=1}$                      &  
\shortstack{$0.69\,e^{-0.05\tau}$ \\ + $10^{-3}$} &
\shortstack{$0.04\,e^{-0.01\tau}$ \\ + $10^{-3}$} &
\shortstack{$2.61\,e^{-0.07\tau}$ \\ + $10^{-3}$} &
\shortstack{$1.59\,e^{-0.06\tau}$ \\ + $10^{-3}$} &
\shortstack{$0.72\,e^{-0.04\tau}$ \\ + $10^{-3}$} & 
\shortstack{$1.94\,e^{-0.05\tau}$ \\ + $10^{-3}$} &
\shortstack{$0.08\,e^{-0.02\tau}$ \\ + $10^{-3}$} &
\shortstack{$1.28\,e^{-0.05\tau}$ \\ + $10^{-3}$} & 
\shortstack{$0.11\,e^{-0.01\tau}$ \\ + $10^{-3}$} & 
\shortstack{$0.09\,e^{-0.01\tau}$ \\ + $10^{-3}$}\\
\bottomrule
\end{tabular}}}
\end{subtable}
\\[1ex]

\begin{subtable}{\textwidth}
\centering
\subfloat[Measured BS-RIS-UE Channel Parameters with RIS]{
\resizebox{\textwidth}{!}{%
\begin{tabular}{l*{10}{c}}
\toprule
Deg.~/~Dist.  & 30$^\circ$~/~3m  & 35$^\circ$~/~3m  & 40$^\circ$~/~3m  & 45$^\circ$~/~3m  & 50$^\circ$~/~3m  & 55$^\circ$~/~3m  & 60$^\circ$~/~3m  & 50$^\circ$~/~5m  & 55$^\circ$~/~5m & 60$^\circ$~/~5m \\
\midrule
$\text{PL}^{\text{BS-RIS-UE}}_{\text{CI}} (\text{dB})$
& 83.1 & 85.9 & 81.3 & 83.7 & 85.9 & 88.4 & 83.6 & 87.8 & 89.8 & 90.6 \\
$\sigma_\text{SF} (\text{dB})$                           &  8.3 & 7.7  &  8.7 & 10.1  &  9 &  7.6 &  9 &  9.3 & 7.6  &  8.1  \\
$K_\text{R} (\text{dB})$                           &  16.2  & 12.9  & 11.3  & 14.6   &  11.6 &  8.4 & 11.9  & 9.9   & 12.4  &  7.8  \\
$\tau_{\text{rms}} (\text{ns})$                           & 11.1  & 14.5  & 15  & 15.6  & 15.1  & 18.1  & 11.5 & 19.5  & 20.7  & 17.1   \\
\midrule
$I_{\text{q}}$                  &  8 & 12  & 8  &  15 &  12 & 9  & 8  &  11 &  8 &  7  \\
$DS_\text{q=1} (\text{ns})$                            & 4.3  &  7.9 &  4.9 & 8  & 8.5  &  9.6 &  5.7 & 14  & 11.5  &  9  \\
$\zeta^2_\text{q=1} (\text{dB})$                     &  3.3 & 2.1 &  3.7 & 1.5  & 4.7  & 5.4  & 3.1  &  4.9 &  6.9 &  3.1  \\
$r^\tau_\text{q=1}$                        & \shortstack{$2.97\,e^{-0.14\tau}$ \\ + $10^{-3}$} &  
\shortstack{$0.08\,e^{-0.05\tau}$ \\ + $10^{-3}$} & \shortstack{$0.05\,e^{-0.03\tau}$ \\ + $10^{-3}$} & 
\shortstack{$0.27\,e^{-0.07\tau}$ \\ + $10^{-3}$} &   
\shortstack{$0.14\,e^{-0.03\tau}$ \\ + $10^{-3}$} &   
\shortstack{$0.3\,e^{-0.02\tau}$ \\ + $10^{-3}$} & 
\shortstack{$0.07\,e^{-0.04\tau}$ \\ + $10^{-3}$} & 
\shortstack{$0.27\,e^{-0.03\tau}$ \\ + $10^{-3}$} & 
\shortstack{$8.91\,e^{-0.15\tau}$ \\ + $10^{-3}$} & 
\shortstack{$1.05\,e^{-0.08\tau}$ \\ + $10^{-3}$}\\
\bottomrule
\end{tabular}}}
\end{subtable}

\end{table*}

\subsection{Delay Spread and K-factor Analysis}


The performance of the \ac{RIS} in the static setup is largely assessed through \ac{RMS} delay spread and K-factor analysis across angles and distances, as shown in Fig.~\ref{fig:ds-kf}. For angular measurements, the delay spread increases at wider angles without \ac{RIS}, as the Rx becomes more exposed to \acp{MPC} coming from the wall located behind the \ac{RIS}. With the \ac{RIS} on, the delay spread decreases and the \ac{vLoS} peak dominates, concentrating energy around the main component (i.e., hardening the channel). For distance measurements, the \ac{RIS} on case shows an increasing delay spread with Rx separation due to added \acp{MPC} from surrounding scatterers, which highlights that the channel hardening effect of the \ac{RIS} drops as the Rx is moved from near-field to far-field. Without the \ac{RIS}, the channel fluctuates without a clear trend, as expected. The K-factor results confirm these trends. The angular sweep without \ac{RIS} shows a narrower distribution around low values (–3.5 dB to 3.5 dB), while the \ac{RIS} on case yields higher values (5.5 dB to 13 dB) due to the stronger \ac{LoS}. The distance sweep shows that the K-factor decreases with RIS-Rx separation as the \ac{vLoS} weakens and multipath grows, consistent with the rising delay spread. In contrast, without the \ac{RIS}, the K-factor remains low and scattered (–7 dB to 2.5 dB).

Clustering analysis also shows that the number of significant delay taps and overall delay spread increase with distance and at wider angles, highlighting how the addition of RIS not only creates the vLoS but also alters the structure of the multipath environment.
 
\footnotetext[3]{The measurement of $X_{n,m}$ XPR values $\left(\mu_{\text{XPR}},\sigma_{\text{XPR}}\right)$ was not performed, because a dual-polarized RIS assumes no change  in the environment, hence their LoS and NLoS values can be taken from Table 7.5-6 of \cite{3GPP1}.}

\subsection{Dynamic UE Measurement Results}
The system-level performance analysis is presented in terms of the throughput in the 2$\times$2 5G NR SU-MIMO system equipped with a 4$\times$4 MIMO smartphone as a \ac{UE}, and the iPerf software to record its data rate at a known transmit power. This setup and the results are shown in Fig. \ref{Throughput}, showing a \ac{CDF} of throughput variation. Two distinctive tests were performed with variable transmit power levels. 

Test 1 evaluated \ac{MF-RIS} sensing indoors with 10 dBm and 30 dBm transmit power, while the UE followed a circular radius 2 m in the near-field of the \ac{MF-RIS} at a constant velocity $\left\|\vec{v}\right\| = 0.1~\text{m/s}$. With \ac{MF-RIS} on, throughput remained consistently high, confirming the accuracy of UE tracking, while disabling \ac{MF-RIS} produced erratic results. With \ac{MF-RIS}, throughput improved by 12.5\%  on average and standard deviation dropped by 74\%, showing strong channel hardening.


Test 2 evaluated the same setup, but with the circular radius increased to 4 m and velocity $\left\|\vec{v}\right\|$ increased to 0.2 m/s. Results mirrored Test 1, with 8.2\% higher average throughput, lower variance, and higher minimum values. Increased RIS-UE distance, representing larger spaces, gave smaller gains (1.85\% average) but still improved minimum throughput and variance by$>$50\%, confirming that \ac{MF-RIS} benefits the quality of \ac{MIMO} channels, even with less focused beams.

\section{Conclusion}
In this work, we derived a \ac{3GPP}-compatible \ac{GBSM} for a \ac{MF-RIS} system and supplemented it with empirical channel measurements. We derived large and small-scale
characteristics of the \ac{MF-RIS}-assisted channel from complex channel responses, including path loss, shadow fading, Rician K-factor, clustering, and \ac{RMS} delay spread. We found that the conventional Rayleigh fading channel significantly deviates from the observed channel response, which is better described by the Weibull distribution over a wide bandwidth or the log-normal distribution over a narrow bandwidth. We also performed a \ac{MIMO} system-level evaluation of the channel throughout, reporting a 74\%  reduction in the throughput variance and a 12.5\% sum-rate improvement in the near-field of the \ac{MF-RIS}, attributed to the increased coherence bandwidth of \ac{MIMO} channels and stronger frequency correlation across subcarriers. 

\bibliographystyle{IEEEtran}
\bibliography{sample}

\end{document}